%% file: main.tex
\documentclass{article}

\usepackage{microtype}
\usepackage{graphicx}
\usepackage{booktabs} 

\usepackage{hyperref}

\usepackage{tabularx}

\usepackage{multirow}
\usepackage{makecell}
\usepackage{hhline}
\usepackage{amsmath}
\usepackage{graphicx}
\usepackage{enumitem}
\usepackage{caption}
\usepackage{subcaption}
\usepackage{amsthm}
\usepackage{comment}

\usepackage[dvipsnames]{xcolor}
\usepackage{listings}
\definecolor{deepblue}{rgb}{0,0,0.5}
\definecolor{deepred}{rgb}{0.6,0,0}
\definecolor{deepgreen}{rgb}{0,0.5,0}
\graphicspath{ {./figures/} }

\usepackage{tikz}

\newcommand{\CompilerName}{RAF}
\newcommand{\CompilerNameWithArticle}{RAF}

\newcommand{\SystemName}{Lancet}
\newcommand{\RegionName}{focus region}

\renewcommand{\paragraph}[1]{\vspace{1mm} \noindent \textbf{#1} \thickspace}

\usepackage[accepted]{mlsys2024}

\mlsystitlerunning{\SystemName{}: Accelerating MoE Training via Whole Graph Computation-Communication Overlapping}

\begin{document}

\twocolumn[
\mlsystitle{\SystemName{}: Accelerating Mixture-of-Experts Training via Whole Graph Computation-Communication Overlapping}



\mlsyssetsymbol{intern_aws}{*}
\mlsyssetsymbol{prev_aws}{\dag}

\begin{mlsysauthorlist}
\mlsysauthor{Chenyu Jiang}{hku,intern_aws}
\mlsysauthor{Ye Tian}{hku,intern_aws}
\mlsysauthor{Zhen Jia}{aws}
\mlsysauthor{Shuai Zheng}{boson,prev_aws}
\mlsysauthor{Chuan Wu}{hku}
\mlsysauthor{Yida Wang}{aws}
\end{mlsysauthorlist}

\mlsysaffiliation{hku}{The University of Hong Kong, Hong Kong}
\mlsysaffiliation{aws}{Amazon Web Services, USA}
\mlsysaffiliation{boson}{Boson AI, USA}

\mlsyscorrespondingauthor{Chenyu Jiang}{jchenyu@connect.hku.hk}
\mlsyskeywords{Machine Learning, MLSys}

\vskip 0.3in

\begin{abstract}
The Mixture-of-Expert (MoE) technique plays a crucial role in expanding the size of DNN model parameters. 
However, it faces the challenge of extended all-to-all communication latency during the training process.
Existing methods attempt to mitigate this issue by overlapping all-to-all with expert computation. 
Yet, these methods frequently fall short of achieving sufficient overlap, consequently restricting the potential for performance enhancements.
In our study, we extend the scope of this challenge by considering overlap at the broader training graph level. 
During the forward pass, we enable non-MoE computations to overlap with all-to-all through careful partitioning and pipelining. In the backward pass, we achieve overlap with all-to-all by scheduling gradient weight computations.
We implement these techniques in \SystemName{}, a system using compiler-based optimization to automatically enhance MoE model training.
Our extensive evaluation reveals that \SystemName{} significantly reduces the time devoted to non-overlapping communication, by as much as 77\%. 
Moreover, it achieves a notable end-to-end speedup of up to 1.3 times when compared to the state-of-the-art solutions.
\end{abstract}
]



\printAffiliationsAndNotice{\textsuperscript{*}Work done during internship at AWS. \textsuperscript{\dag}Work done while at AWS.}  

\input{sections/introduction}

\input{sections/background_motivation}

\input{sections/overview}

\input{sections/dw_schedule}

\input{sections/op_partition}

\input{sections/implementation}

\input{sections/evaluation}

\bibliography{references}
\bibliographystyle{mlsys2024}



\end{document}

%% file: sections/introduction.tex
\section{Introduction}

Recent research has prompted a continuous trend of constructing larger DNN models across application domains. 
However, directly adopting wider or deeper network architecture typically leads to a proportional increase in computation.
In contrast, Mixture of Experts (MoE)~\cite{ShazeerMMDLHD17MoE, lepikhin2020gshard} has the ability to increase the parameter size without escalating the total computation.
It has enabled scaling model parameters to the trillion-level~\cite{yang2021m6, lin2021m610t,fedus2022switch, nie2022hetumoe}, showcasing the superior performance compared to dense counterparts~\cite{fedus2022switch, hwang2023tutel, rasley2020deepspeed}.

\begin{figure}[t]
    \centering
    \includegraphics[width=\linewidth]{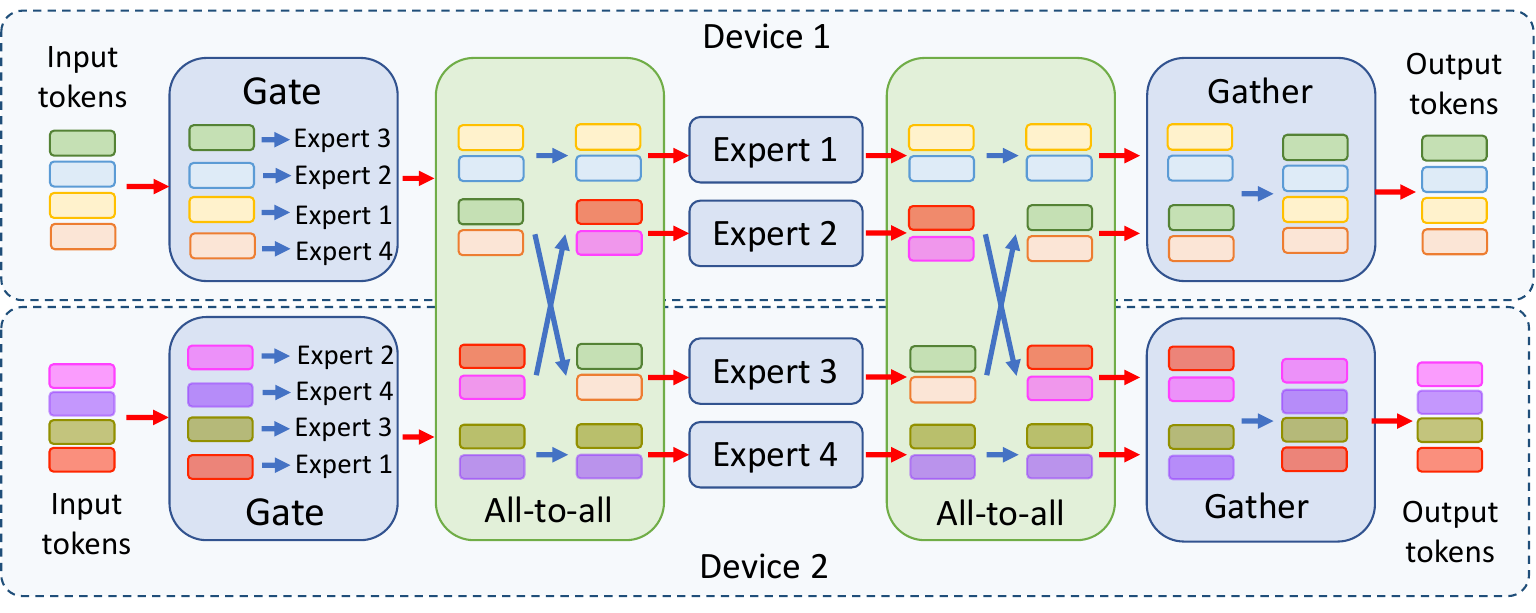}
    \vspace{-7mm}
    \caption{
    An example MoE layer with 4 experts scattered on 2 devices. Assume top-1 gating is used. Blue (green) boxes represent computation (communication) operators. Data dependency between operators are highlighted by red arrows. The \textit{Gate} assigns each input token to an expert. \textit{All-to-all}s fetch expert input/output from other devices. \textit{Gather} restores the received tokens back to their original order, matching the input sequence.
    }
    \label{fig:moe_and_sharding}
\end{figure}

Efficient parallelization of MoE models requires assigning distinct experts to separate accelerator devices~\cite{lepikhin2020gshard}.
Yet, distributing input samples to these scattered experts demands resource-intensive all-to-all communication (Fig.~\ref{fig:moe_and_sharding}).
High communication volume in all-to-all operations significantly hampers the training speed of MoE models (up to 40\% of training time).

For non-MoE models, communication scheduling~\cite{Jayarajan19priority,peng2019generic} is an effective way to overlap the communication (for synchronizing model parameters) and backward propagation.
However, they are inapplicable for MoE models, which have a direct data dependency between all-to-all and other computations (experts and non-MoE computation like self-attention), as in Fig.~\ref{fig:moe_and_sharding}.
For MoE models, existing studies~\cite{hwang2023tutel,he2022fastermoe,wang2022overlap,li2023automated} focused on alleviating this problem by partitioning operators into finer-grained ones and overlapping communication with computation between different partitions.
Nonetheless, their \RegionName{} is limited to encompass only the all-to-all communication and expert computation. 
In this paper, we define the \RegionName{} as the subset of operators within the training graph responsible for concurrent (overlapping) computation and communication.
We have observed that the all-to-all communication time is usually much longer than expert computation time, thus the overall execution time is still bounded by the all-to-all communication despite overlapping (Fig.~\ref{fig:motivation_exec_time_breakdown}).
The small \RegionName{} considered in existing works limits the overlapping possibilities and thus results in the sub-optimal performance.

In this paper, we extend the \RegionName{} to the whole training graph and identify two more types of operators to overlap: 1) weight gradient computation in backward pass, which does not depend on all-to-all communication and thus is able to overlap with it directly. 2) non-MoE model computation in forward pass, which has dependency with all-to-all but can perform overlapping by properly partitioning.
However, extending the \RegionName{} also raises new challenges: 
1) Extending overlapping to non-MoE computation requires partitioning along batch dimension. A direct partition may cause mathematical in-equivalency since the routing decision of many gating methods can be affected by batch size.
2) Partitioning introduces more smaller operators, thereby incurring GPU kernel launching overhead and under-utilization of streaming multiprocessors.
Over-partitioning computations can lead to excessive overhead, negating the benefits of overlapping. 
Conversely, insufficient partitioning hinders the full utilization of potential overlap with all-to-all communication.
Additionally, gating methods limit the types of operators that can be partitioned. 
Hence, extending the \RegionName{} introduces complexity in establishing the best partitioning range, which refers to the number of computation operators preceding and succeeding an all-to-all communication operator that need to be partitioned (and overlapped with the all-to-all communication).

To overcome those challenges, we propose \SystemName{}, a system designed to enhance the throughput of MoE model training by considering the entire training graph as \RegionName{}.
\SystemName{} leverages a compiler-based approach, providing us with increased flexibility for controlling operator partitioning and scheduling. Distinct mechanisms are applied for the forward and backward passes during training. In the forward pass, where nearly all computations rely on all-to-all dependencies, it becomes necessary to partition both computation and all-to-all operators to achieve efficient overlaps. In the backward pass, we employ scheduling to ensure the weight gradient computation overlaps with all-to-all operations.
The rationale behind this approach lies in the backward pass, where there are an ample number of weight gradient computation operators that can be scheduled to enable near-complete overlap with all-to-all operations. As a result, there is no need to explore partitioning solutions, as is required in the forward pass.
The method we designed to overlap all-to-all with entire training graph does not conflict with non-MoE model communication scheduling strategies~\cite{Jayarajan19priority, peng2019generic}.
And all transformations (scheduling and partitioning) maintain mathematical equivalence (i.e., the model accuracy remains unaffected by the optimizations) and can be kept transparent to users.

In summary, our contributions include:

$\triangleright$ For the first time, we expand the \RegionName{} to encompass the entire training graph, mitigating the prolonged all-to-all communication's impact on MoE model training.
This extension enables us to discover new operators that can be overlapped with all-to-all communication.

$\triangleright$ We adopt a greedy algorithm to schedule each weight gradient computation operator to overlap with the appropriate all-to-all.

$\triangleright$ We devise a partitioning scheme for MoE layers that allows for the extension of partitioning to non-MoE computations while maintaining mathematical equivalency.

$\triangleright$  We apply a dynamic programming based algorithm to identify the optimal range of non-MoE computation for partitioning and overlapping.

Comprehensive evaluations demonstrate that \SystemName{} can decrease non-overlapping communication time by as much as 77\% and deliver an up to 1.3x end-to-end speedup when compared to state-of-the-art solutions, including DeepSpeed~\cite{rasley2020deepspeed} and Tutel~\cite{hwang2023tutel}.

%% file: sections/background_motivation.tex
\section{Background and Motivation}
\vspace{-1mm}

\subsection{Mixture of Experts (MoE)}
Most MoE models~\cite{ShazeerMMDLHD17MoE, lepikhin2020gshard} replace the feed-forward module in every two Transformer layers with multiple independent sub-networks (experts), each activated by a subset of input data.
Different experts are usually placed on distinct devices for efficient parallelization.
In this work, we assume non-MoE parts of the model are replicated across the devices while receiving different partitions of the training data (i.e., data parallelism).
The assignment (routing) of inputs to experts is decided by a gating function at runtime.

\vspace{-3mm}
\paragraph{Expert-parallelism}
Once the gating function determines expert assignments, all-to-all communication transmits inputs to the respective devices.
After expert processing, another all-to-all operation sends their output back to the original devices.
Due to dynamic expert assignment at runtime, token distribution among experts varies. 
To maintain static tensor shapes (essential for certain frameworks/hardware like XLA/TPU~\cite{lepikhin2020gshard}) and ensure balanced computation across experts, a common approach is to restrict the maximum tokens assigned to each expert (expert capacity, $C$) on each device~\cite{lepikhin2020gshard, fedus2022switch}. 
Any excess tokens assigned to an expert are discarded, while experts receiving fewer than $C$ tokens are zero padded.

\vspace{-3mm}
\paragraph{Routing algorithms}
Routing is often computed by assigning a gating score for each expert using a trainable linear layer, and choosing $k$ experts with highest scores (top-k routing)~\cite{lepikhin2020gshard, fedus2022switch}.
Recent works also propose other routing methods, such as hash-based~\cite{roller2021hash} or random expert assignment~\cite{zuo2022taming, chen2023sparse} and expert-choice routing~\cite{zhou2022expertchoice} (experts choose top-k tokens with highest scores).
Routing algorithms significantly influence MoE model training, impacting expert balance~\cite{zhou2022expertchoice}, communication costs~\cite{he2022fastermoe} and more.
The upcoming sections also demonstrate how routing algorithms affect feasible optimizations.

\subsection{Overlapping all-to-all and experts}

\begin{figure}[t]
     \centering
     \includegraphics[width=\linewidth]{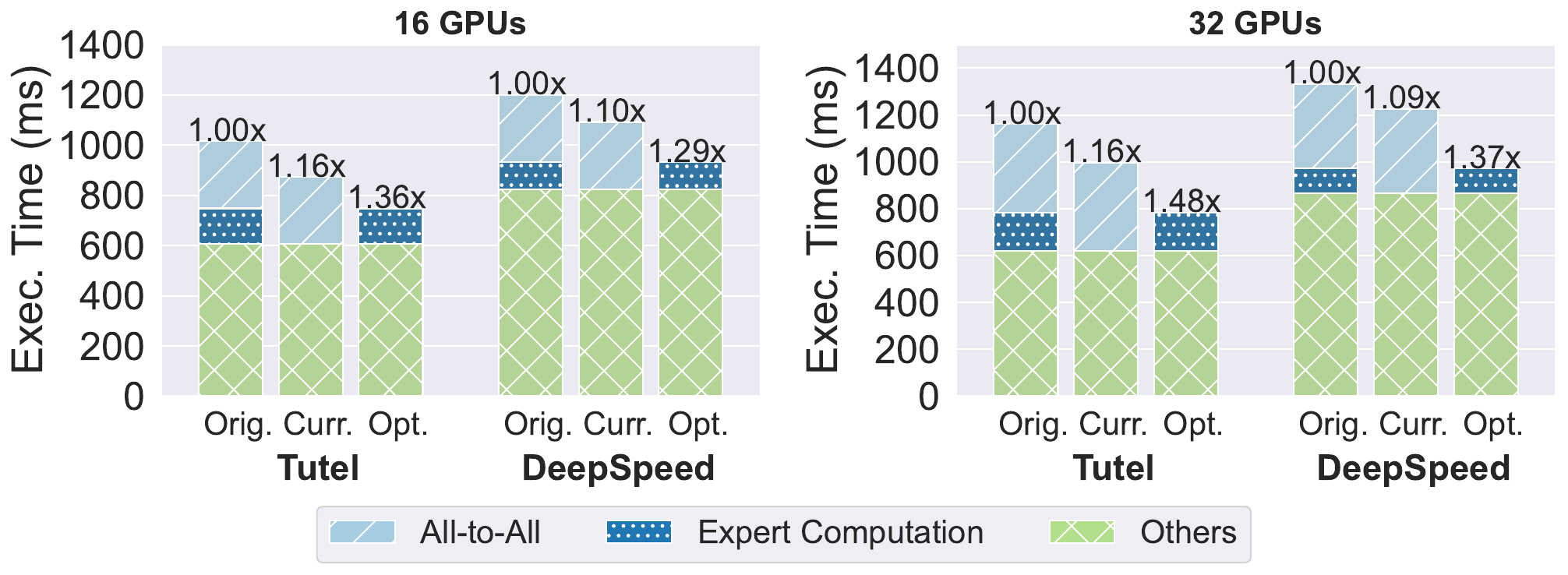}
    \caption{Breakdown of execution time when running a GPT-2 model with MoE layers using Tutel and DeepSpeed on Amazon EC2 p3dn instances. \textit{Orig.}: unoptimized execution time. \textit{Curr.}: performance upper-bound when optimized using current overlapping methods (expert computation completely hidden by all-to-all). \textit{Opt.}: ideal execution time (all-to-all fully overlapped by computation).}
    \label{fig:motivation_exec_time_breakdown}
\end{figure}

Existing techniques aim to mitigate long latency all-to-all impacts on training by overlapping it with expert computation~\cite{hwang2023tutel, he2022fastermoe}. 
This involves partitioning all-to-all and experts along the capacity dimension and forming a communication-computation pipeline with (only) all-to-all and experts (Fig.~\ref{fig:current_partition}). 
As shown in Fig.~\ref{fig:motivation_exec_time_breakdown}, we often observe the all-to-all time significantly surpasses that of the experts (up to 3.36x).
Therefore, these techniques can only conceal the execution time of experts, while the bottleneck execution time for all-to-all communication remains unaffected.

\begin{figure}[t]
     \centering
     \begin{subfigure}[]{\columnwidth}
         \centering
         \includegraphics[width=\linewidth]{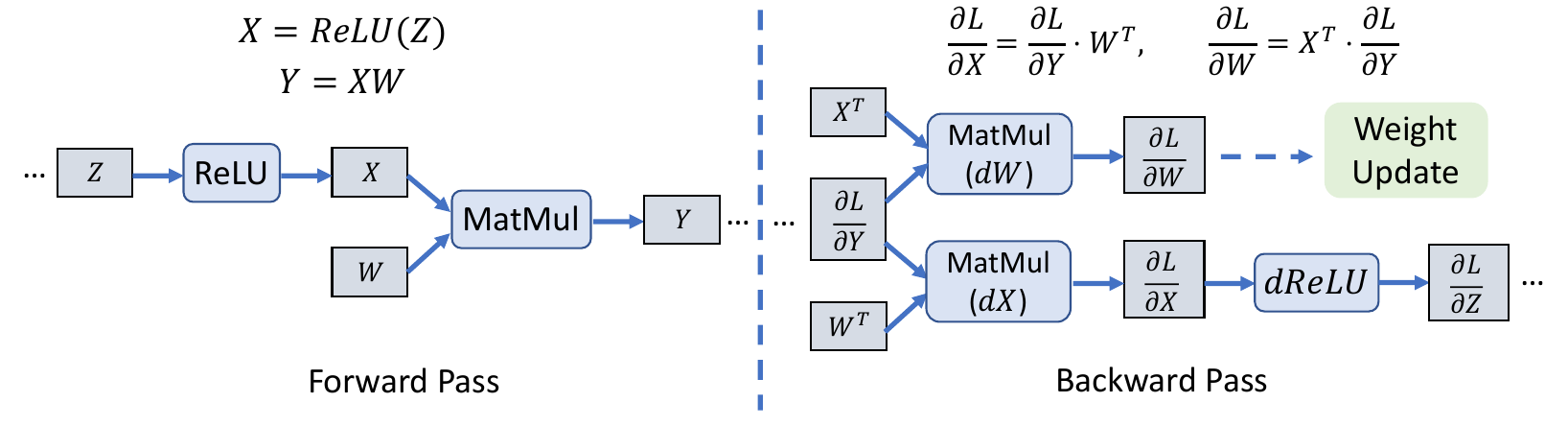}
         \vspace{-5mm}
         \caption{Forward and backward pass of $Z=ReLU(X);Y=ZW$. $dX$: activation gradient computation; $dW$: weight gradient computation. Gray block: tensors; Blue block: operators. Note that $dReLU$ (and further back-propagation of $\frac{\partial L}{\partial Z}$, if needed) does not depend on $dW$.}
         \label{fig:example_dw}
     \end{subfigure}
     \hfill
     \begin{subfigure}[]{\columnwidth}
         \centering
         \includegraphics[width=\linewidth]{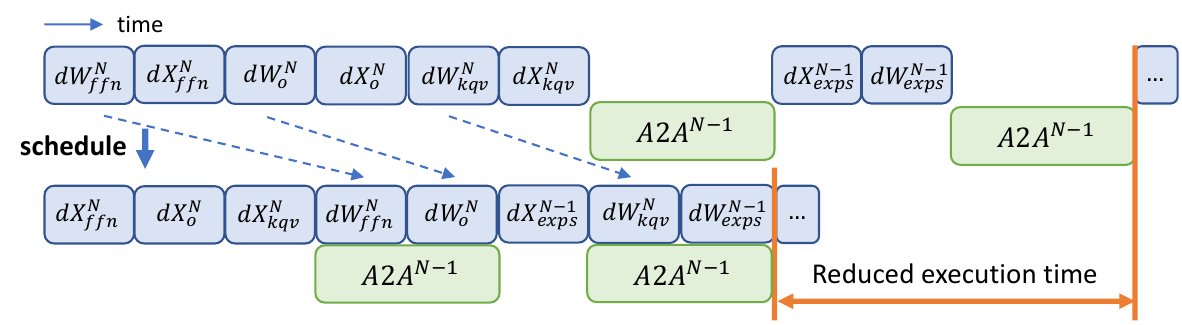}
         \vspace{-5mm}
         \caption{Overlapping all-to-all communication by scheduling weight gradient computation ($dW$). Superscripts on operators indicate the layer number, with the first layer after the MoE layer (during forward) numbered layer $N$. Subscripts indicates the type of operators: \texttt{ffn}: non-expert feed-forward; \texttt{o}: output projection in self-attention; \texttt{kqv}: key, query and value projection; \texttt{exps}: experts. Other operators are ignored. While the figure shows overlapping all-to-all in layer $N-1$ with $dW$s in layer $N$, in general the all-to-all can be overlapped with any $dW$s in layer $N+k, k\geq 0$.}
         \label{fig:example_scheduling}
     \end{subfigure}
    \vspace{-3mm}
    \caption{Scheduling weight gradient computation to overlap with all-to-all.}
    \label{fig:schedule_dW}
\end{figure}

\subsection{Opportunities and Challenges}

By extending the \RegionName{} to the whole training graph, we identify more opportunities that can overlap with all-to-all communications.

\paragraph{Opportunity 1: Weight gradient computation.}
Computation of the weight gradient in layer $N$, which is essential for updating model parameters, is independent of all-to-all communication of previous layers $N-1,N-2,\ldots,1$ in the backward pass. Consequently, it can be scheduled to overlap with all-to-all, allowing for flexibility in optimizing the training time (Fig.~\ref{fig:schedule_dW}).

\begin{figure}[t]
     \centering
     \begin{subfigure}[]{\columnwidth}
         \centering
         \includegraphics[width=\linewidth]{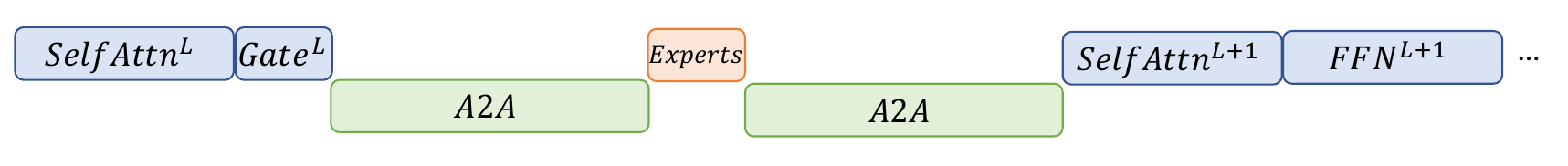}
         \vspace{-5mm}
         \caption{No overlapping}
         \label{fig:no_partition}
     \end{subfigure}
     \hfill
     \begin{subfigure}[]{\columnwidth}
         \centering
         \includegraphics[width=\linewidth]{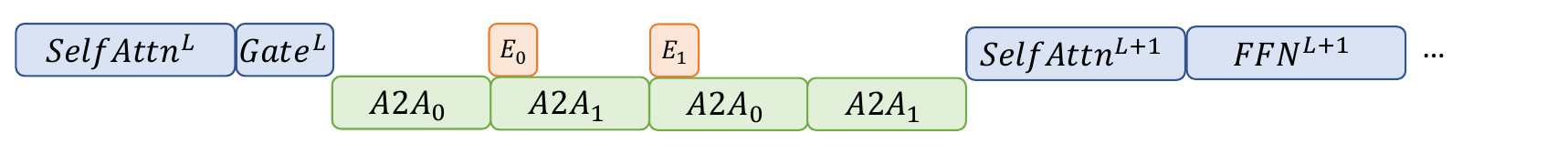}
         \vspace{-5mm}
         \caption{Overlapping all-to-all and expert computation}
         \label{fig:current_partition}
     \end{subfigure}
     \hfill
     \begin{subfigure}[]{\columnwidth}
         \centering
         \includegraphics[width=\linewidth]{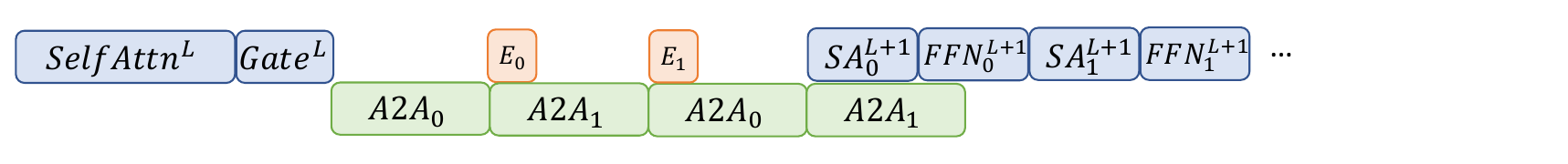}
         \vspace{-5mm}
         \caption{Overlap all-to-all, experts and the non-MoE computation after the current MoE layer.}
         \label{fig:part_after_gate}
     \end{subfigure}
     \hfill
      \begin{subfigure}[]{\columnwidth}
         \centering
         \includegraphics[width=\linewidth]{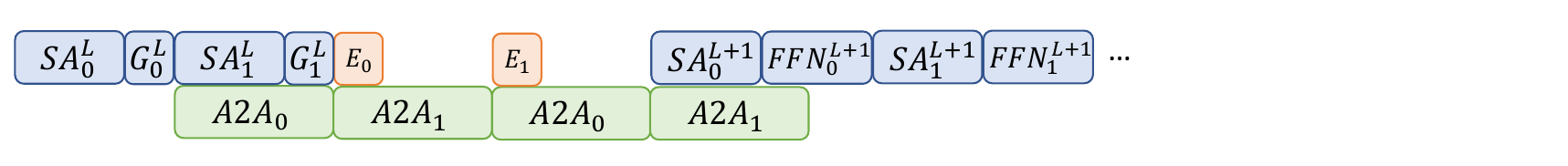}
         \vspace{-5mm}
         \caption{Also overlap non-MoE computation before the current MoE layer.}
         \label{fig:part_all}
     \end{subfigure}
    \vspace{-3mm}
    \caption{Performance gain of different overlapping types.}
    \label{fig:partition_ranges}
    \vspace{-2pt}
\end{figure}

\paragraph{Opportunity 2: Non-MoE computation.}
In existing systems (e.g.,~\cite{hwang2023tutel, he2022fastermoe}), computation before (e.g., self-attention) and after (e.g., the following Transformer layer) the MoE layer does not overlap with all-to-all communication since their limited \RegionName{}.
Nevertheless, if we partition non-MoE computations and integrate them into the computation-communication pipeline, we can create additional opportunities to overlap operations with the all-to-all communication (Fig.~\ref{fig:part_after_gate}, \ref{fig:part_all}).

\begin{figure}[t]
     \centering
     \begin{subfigure}[]{\columnwidth}
         \centering
         \includegraphics[width=\linewidth]{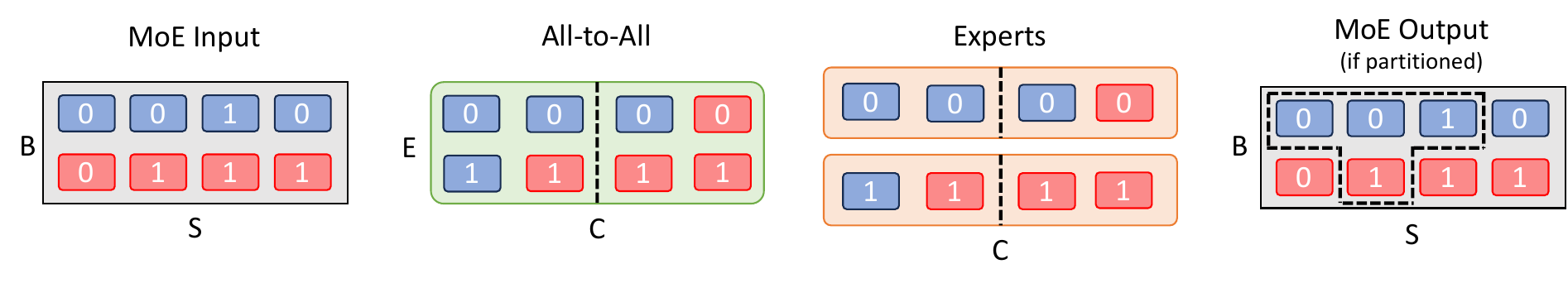}
         \vspace{-5mm}
         \caption{Operator partition dimensions of Tutel. All-to-all and experts are partitioned at capacity dimension; tokens belong to different partitions appear at irregular locations in the output of MoE layer.}
         \label{fig:current_part_tokens}
     \end{subfigure}
      \begin{subfigure}[]{\columnwidth}
         \centering
         \includegraphics[width=\linewidth]{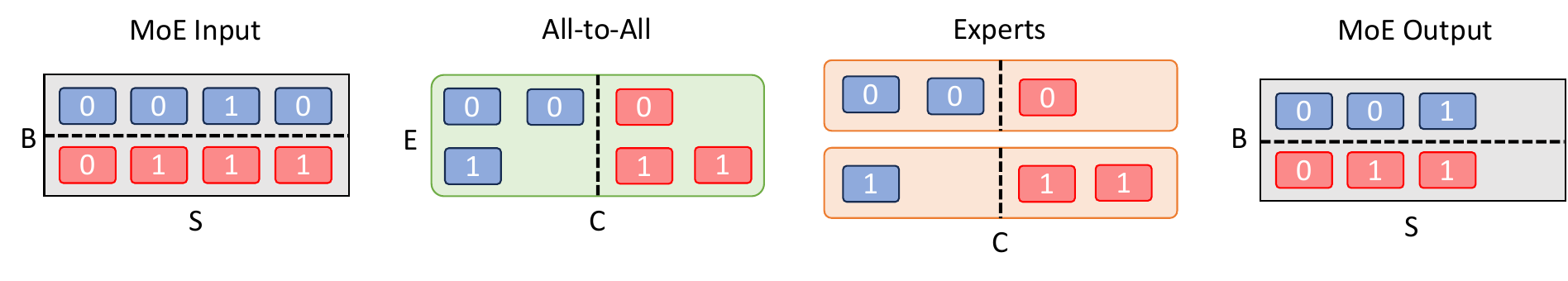}
         \vspace{-5mm}
         \caption{Direct micro-batching. All-to-all and experts capacity also drops proportionally, causing extra token dropping.}
         \label{fig:direct_slice_tokens}
     \end{subfigure}
     \hfill
     \begin{subfigure}[]{\columnwidth}
         \centering
         \includegraphics[width=\linewidth]{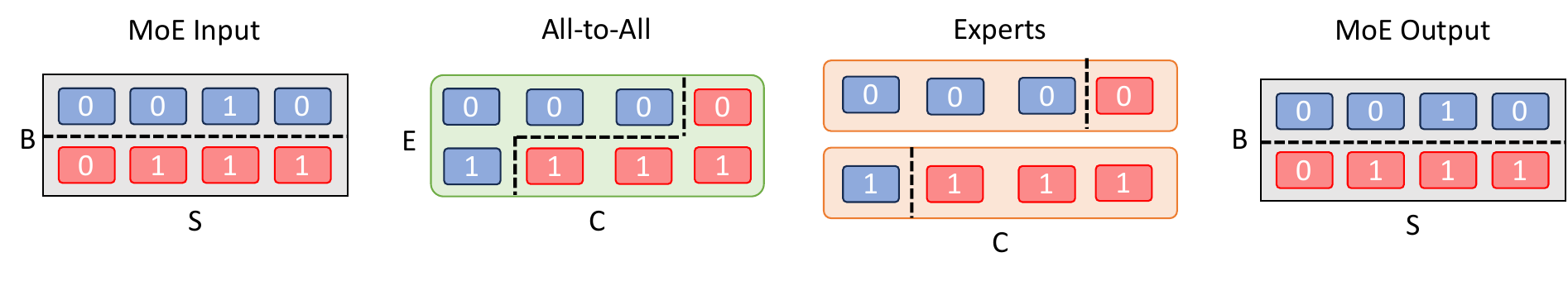}
         \vspace{-5mm}
         \caption{Micro-batching with irregular expert capacity. All-to-all and experts are irregularly partitioned, while MoE inputs and outputs are partitioned at batch dimension (facilitating further pipelining).}
         \label{fig:proposed_partition}
     \end{subfigure}
    \caption{Operator partitioning scheme in an MoE layer. Number in each token shows their assigned expert. Tokens of the same color belong to the same sequence.}
    \label{fig:partition_ranges_tokens}
    \vspace{-4mm}
\end{figure}

\paragraph{Challenge 1: How to perform mathematically equivalent partition.}
Consider feeding a tensor with dimension $B \times S$ ($B$ as batch size, $S$ as sequence length) into an MoE layer (in ~Fig.~\ref{fig:partition_ranges_tokens}).
The tokens are re-arranged according to their target experts, undergoing an all-to-all with shape $E \times C$ ($E$ as the total number of experts, $C$ as the expert capacity) for distributing to the corresponding device.
Each expert processes the $C$ received tokens, followed by a reciprocal all-to-all (not shown in the figure).
Reverting tokens to their original order yields the MoE layer's output.
The existing methods all partition the all-to-all and experts at the capacity dimension ($C$), and thus the tokens in same partition appear in irregular locations in the re-arranged MoE output, e.g., belong to different sequences across the batch (Fig.~\ref{fig:current_part_tokens}).
Therefore, the following computation must wait until all partitions finish execution, interrupting the pipeline.

In order to overlap forward pass non-MoE computation with all-to-all, the MoE input and output must be partitioned along the batch dimension.
However, directly partitioning the input (micro-batching) may result in extra token dropping since the expert capacity also drops accordingly. For example, consider a input batch (with corresponding expert capacity $C$) partitioned into two micro-batches, each processed with expert capacity $\frac{1}{2}C$. Assume the first micro-batch contains $\frac{3}{4}C$ tokens for an expert, and the second contains $\frac{1}{4}C$ tokens for that expert. If not partitioned, then all tokens can fit into expert capacity $C$ thus no token will be dropped. When directly partitioned, $\frac{1}{4}C$ tokens will be dropped from the first micro-batch since it now only has expert capacity $\frac{1}{2}C$(Fig.~\ref{fig:direct_slice_tokens}).
Such change in mathematical equivalency is undesirable as it may affect model performance.

To avoid this effect, we implement special gating operators that pass capacity information between partitions (e.g., when the first partition (micro-batch) uses $\frac{3}{4}C$ capacity, the second partition will adjust its remaining capacity to $\frac{1}{4}C$), preserving the exact token-to-expert mapping and token dropping as the un-partitioned case.
This however implies that any partition can send any amount of token (ranging from 0 to $C$) to an expert (while tokens sent from all partitions add up to $C$) (Fig.~\ref{fig:proposed_partition}). We implement irregular-shaped all-to-all to efficiently handle such a dynamic communication pattern (details discussed in Sec.~\ref{sec:implementation}).

\begin{figure}[t]
     \centering
      \begin{subfigure}[]{0.49\columnwidth}
         \centering
         \includegraphics[width=\linewidth]{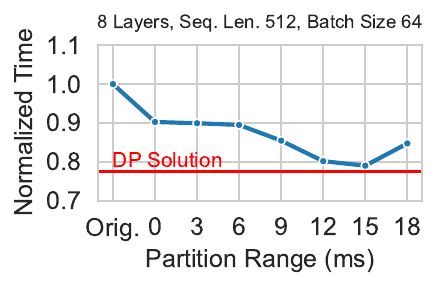}
         \vspace{-5mm}
         \caption{Less layers, large batch size}
         \label{fig:part_range_wide}
     \end{subfigure}
     \hfill
    \begin{subfigure}[]{0.49\columnwidth}
         \centering
         \includegraphics[width=\linewidth]{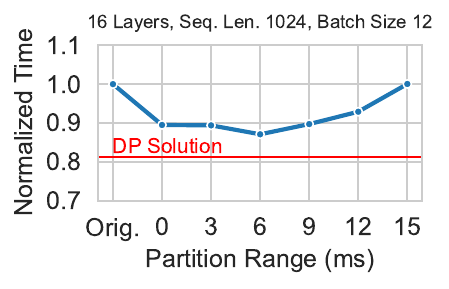}
         \vspace{-5mm}
         \caption{More layers, small batch size}
         \label{fig:part_range_long}
     \end{subfigure}

    \caption{Effect of partition range on GPT-2 MoE model forward time on 16 A100 GPUs (32 experts). X axis shows how many ops (measured in their execution time) before and after the MoE layer is included in the partition. \textit{Orig.}: no partitioning. \textit{0}: only partition all-to-all and experts (as in Tutel).}
    \label{fig:part_range_selection}
\end{figure}

\paragraph{Challenge 2: How to determine the optimal partition range for non-MoE operators.}
GPU kernel launches involve startup overhead~\cite{rotem2018glow}, which occurs with each launch.
Partitioned computation operators deal with smaller input tensors, potentially leading to GPU core under-utilization.
Similarly, smaller communication operators might not fully utilize network bandwidth.
So it is not always optimal to partition the entire Transformer layer before and after MoE layer,
which may increase training time due to partition overheads. Fig.~\ref{fig:part_range_selection} shows this phenomena.
The optimal partition range (a set of computation ops around all-to-all) depends on model specification, input size, the underlying computation power and also network bandwidth. 
Our extension of the \RegionName{} to whole training graph makes the decision more challenging. 

Furthermore, the gating methods limit the partitioning opportunities.
Some gating methods assign target experts based on the information calculated over the entire batch of tokens.
For example, Batch-prioritized Routing~\cite{riquelme2021bpr} sorts tokens in a batch first by their ``importance score" (the sum of top-k largest gating scores) and then assigns tokens to experts. So tokens with lower scores would be dropped first.
Splitting along batch dimension would thus cause differences in token dropping. For such gating methods, we can only extend partitioning after the MoE layer (Fig.~\ref{fig:part_after_gate}). For other gating methods whose expert assignment can be decided from partial batches (e.g., Switch~\cite{fedus2022switch} or Random~\cite{zuo2022taming, chen2023sparse} gating), we can extend partitioning to both after and before the MoE layer (Fig.~\ref{fig:part_all}).

%% file: sections/overview.tex
 \section{\SystemName{} Overview}
\label{sec:overview}

\begin{figure}[t]
    \centering
    \includegraphics[width=0.7\columnwidth]{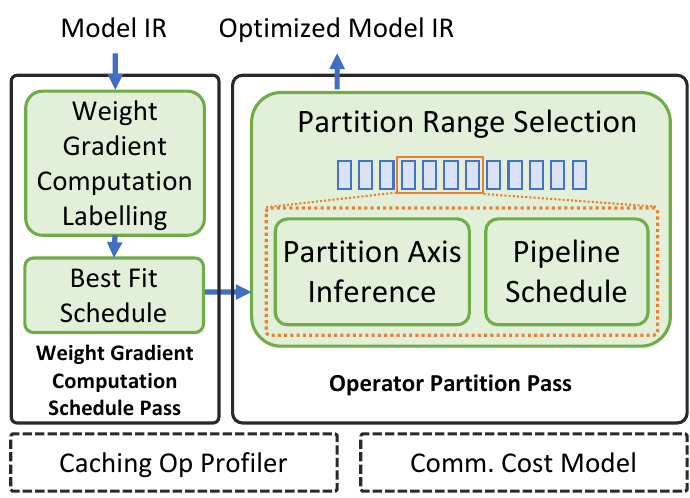}
    \vspace{-3mm}
    \caption{Overview of \SystemName{} modules.
    }
    \label{fig:overview}
\end{figure}

To address the opportunities and challenges mentioned above, we devised \SystemName{}, a compiler-based solution designed to optimize MoE model training.
The key advantage of this approach is the explicit extraction of a model's computation and communication through 
compilers' intermediate representation (IR), granting precise control over operator execution and simplifying the implementation of
weight gradient computation scheduling and the partition of operators.

\SystemName{} adopts compiler passes to enhance MoE model training and resolve associated issues. At a higher level, it encompasses two primary optimization passes: weight gradient computation scheduling and operator partitioning, modifying the backward and forward pass of the model, respectively. Fig.~\ref{fig:overview} gives an overview of \SystemName{}.

1) \textbf{Weight Gradient Computation Schedule Pass (\S\ref{sec:dw_schedule_pass})} takes the model IR as input, which is a sequence of instructions, and re-orders the instructions corresponding to weight gradient computation operators to overlap with all-to-alls during backward propagation.
Dependency analysis is first performed to identify the weight gradient computation instructions that can be overlapped with each all-to-all (\S\ref{sec:dw_labelling}).
Then for each all-to-all op, we employ a best-fit greedy algorithm to choose a set of weight gradient computation ops with comparable total execution time to maximize overlap (\S\ref{sec:dw_scheduling}).

2) \textbf{Operator Partition Pass (\S\ref{sec:operator_partition})} receives the IR with weight gradient computation scheduled and further optimizes the all-to-alls in the forward pass through partitioning and pipelining.
A dynamic programming algorithm is employed to find the optimal partition range for non-MoE ops (\S\ref{sec:pipeline_range_selection}).
During this process, a partition axis inferencer (\S\ref{sec:partition_axis_inference}) employs a constraint programming algorithm to deduce the partition axis for each instruction, facilitating partitioning of IR.
Then, a pipeline scheduler (\S\ref{sec:pipeline_schedule}) estimates the cost of the resulting computation-communication pipeline, guiding the dynamic programming algorithm.

These optimizations are supported by a {\bf Caching Op Profiler}, which profiles and caches the execution time of all ops in the model IR.
Profiling is done once for each (partitioned) operation with the same shape; the cached execution time can be subsequently reused.
Communication costs (e.g., partitioned all-to-all) are estimated by a {\bf Communication Cost Model}.
The communication cost model is built by profiling communication operations across various input sizes (e.g., 1KB, 2KB, 4KB,$\cdots$, up to the maximum possible communication used in models), and the cost is linearly interpolated among these points.
Since Lancet uses irregular-shaped all-to-alls (Fig.~\ref{fig:irregular_implementations}), their execution time depend on the combination of actual amount of data to be communicated, which is not known at compilation time.
Therefore, we resort to a static-shape approximation: the cost of an $n$-partitioned all-to-all with original capacity $C$ is obtained by querying the profiled (uniform-shaped) cost model at capacity $C/n$.
We observe that such approximation suffices to produce a good prediction of overall iteration time during our experiments (Fig.~\ref{fig:preduction_accuracy}).

%% file: sections/dw_schedule.tex
\section{Weight Gradient Computation Schedule Pass}
\label{sec:dw_schedule_pass}

The weight gradient computation schedule pass takes the model IR describing the training iteration as input and re-orders the instructions to overlap weight gradient computation with all-to-alls.
The IR is represented as a sequence of instructions $\mathcal{I}=[I_1, I_2,\cdots,I_N]$;
each instruction is characterized by its input tensors $\mathbf{x}$, output tensors $\mathbf{y}$, and operator $f$: $I_n = (\mathbf{x^n}, \mathbf{y^n}, f^n)$, representing the operation $y^n_1,y^n_2,\cdots,y^n_{|\mathbf{y^n}|} = f^n(x^n_1, x^n_2,\cdots,x^n_{|\mathbf{x^n}|})$.

\subsection{Weight Gradient Computation Labelling}
\label{sec:dw_labelling}
Due to the fine-grained nature of instructions, identifying weight gradient computation instructions becomes challenging.
While there is no direct dependency between weight gradient computation and all-to-alls, the weight gradient computations must adhere to the constraints imposed by the chain rule, which imposes scheduling restrictions.
Therefore, we first identify the set of weight gradient computation instructions that can be overlapped with each all-to-all by analyzing instruction dependencies. Consider a dependency graph $\mathbf{G}=(\mathcal{I}, \mathcal{E})$, where each directed edge $E_{i,j}$ asserts that $I_j$ depends on $I_i$, i.e., $I_j$ consumes the output of $I_i$ thus must be executed after it. Then, a weight gradient computation instruction $I_i$ can be overlapped with an all-to-all instruction $I_a$ if and only if there is no directed path between $I_i$ and $I_a$ in $\mathbf{G}$. Such paths can be discovered by a simple Depth- or Breadth-first Search algorithm. For each all-to-all instruction $I_a$, we compute the set of instructions that can overlap with it as $\mathbf{W}^{I_a}$, which is used in the scheduling algorithm.

\subsection{Weight Gradient Computation Scheduling}
\label{sec:dw_scheduling}

We then optimize the scheduling of labelled weight gradient computation operations to minimize the overall training time.
Determining the schedule of weight gradient computation is equivalent to deciding an assignment of each weight gradient computation operator to an all-to-all with which it will overlap. Let $\mathcal{I}^{W}$ be the sub-sequence of $\mathcal{I}$ containing all weight gradient computation instructions, and $\mathcal{I}^{a}$ be the sub-sequence containing all all-to-alls. Let variable $x_{i,j} = 1$ if $I^{W}_i$ (the $i$th weight gradient computation instruction) is assigned to $I^{a}_j$ (the $j$th all-to-all), and otherwise $x_{i,j} =0$. The execution time of $I^{W}_i$ ($I^{a}_i$) is $t^{W}_i$ ($t^{a}_i$). Then maximization of total overlapped all-to-all execution time can be formulated as the following integer program:
\begin{align*}
\max_{\mathbf{x}} \quad & \sum_{j=1}^{|\mathcal{I}^a|}{\min\{t^{a}_j, \sum_{i=1}^{|\mathcal{I}^{W}|}{t^{W}_i \cdot x_{i,j}}\}} \\
\textrm{s.t.} \quad & \sum_{j=1}^{|\mathcal{I}^a|}{x_{i,j}} \leq 1, \quad\forall\ i\in [1, |\mathcal{I}^{W}|] \tag{1}\label{eq:sch_constraint1}\\
  & x_{i,j} = 0, \quad\quad\quad\forall\ I^{W}_i \notin \mathbf{W}^{I^a_{j}} \tag{2}\label{eq:sch_constraint2}
\end{align*}
The $\min\{t^{a}_j, \sum_{i=1}^{|\mathcal{I}^{W}|}{t^{W}_i \cdot x_{i,j}}\}$ gives the amount of overlapped time in each all-to-all. Constraint \eqref{eq:sch_constraint1} states that each weight gradient computation instruction can only be used to overlap with at most one all-to-all. 
\eqref{eq:sch_constraint2} restricts the assignment based on instruction dependency calculated during weight gradient computation labelling. 

Such a problem is a generalized assignment problem (GAP) with non-linear objective and additional constraints \eqref{eq:sch_constraint2}. Since GAP is already known to be NP-hard~\cite{martello1990knapsack}, we resort to a greedy heuristic. We sequentially iterate through $\mathcal{I}^{a}$: for each $I^a_i$, weight gradient computation instructions are greedily chosen from $\mathbf{W}^{I^a_i}$, that are not already used to overlap with other all-to-alls and minimize the absolute difference between the all-to-all execution time and sum of all weight gradient computation to be overlapped with it. We proceed to the next all-to-all when the current one is fully overlapped.

After deciding the assignment of weight gradient computation, we reorder the instructions, placing them right after their overlapping all-to-all instructions. This ensures the weight gradient computation start execution immediately following the launch of all-to-all communication. Alg.~\ref{algo:dw_schedule} presents the 
entire weight gradient computation scheduling process.

\renewcommand{\algorithmicrequire}{\textbf{Input:}}
\renewcommand{\algorithmicensure}{\textbf{Output:}}
\newcommand{\LineComment}[1]{\Statex \hfill/* \textit{#1} */}

\begin{algorithm}[!t]
\caption{Weight Gradient Computation Schedule Pass}\label{algo:dw_schedule}
\begin{algorithmic}[1]
\REQUIRE{$\mathcal{I}$ - a sequence of instructions}
\ENSURE{$\mathcal{I}'$ - scheduled instructions}

\STATE $\mathbf{G} \gets$ CreateDependencyGraph($\mathcal{I}$)
\STATE \small{\textcolor{blue}{/* Weight gradient computation labelling */}}
\STATE $\mathcal{I}^a \gets [I_i\in\mathcal{I}| f^i\text{ is all-to-all}]$\;
\STATE {$\mathbf{W}^{I_i}\gets \{\}$ \textbf{for each} $I_i \in \mathcal{I}^a$}
\FOR{$I_i \in \mathcal{I}^a$, $I_j \in \mathcal{I}, I_i\neq I_j$}
    \IF{no directed path between $I_i$ and $I_j$}
        \STATE $\mathbf{W}^{I_i}$.insert($I_j$)
    \ENDIF
\ENDFOR
\STATE \small{\textcolor{blue}{/* Weight gradient computation scheduling */}}
\STATE $\mathbf{t}^a, \mathbf{t}^{W} \gets$ GetInstrExecTime($\mathcal{I}$)
\STATE $\mathbf{W}^{\text{used}} \gets \{\}$
\STATE $\mathbf{Asg}\gets \{\}$ \small{\textcolor{blue}{ /* map recording the assignment results */}}
\FOR{$i \in |\mathcal{I}^a|$}
    \STATE $t_{u}\gets \mathbf{t}^a_i$ \small{\textcolor{blue}{ /* unoverlapped time of all-to-all i */}}
    \WHILE{$t_{u}>0$ and $\mathbf{W}^{I^a_i}\cap (\mathcal{I}-\mathbf{W}^{\text{used}}) \neq \emptyset$}
        \STATE \small{\textcolor{blue}{/* Find available instr that best matches $t_{u}$ */}}
        \STATE $j_{\text{min}} \gets \text{argmin}_{j}\{\lvert t_{u} - t^{W}_j \rvert\big{|}I^{W}_j \in \mathbf{W}^{I^a_i}, I^{W}_j \notin \mathbf{W}^{\text{used}}\}$
        \STATE $t_{u} \gets t_{u} - t^{W}_{j_{\text{min}}}$
        \STATE $\mathbf{W}^{\text{used}}$.insert($I^{W}_{j_{\text{min}}}$)
        \STATE $\mathbf{Asg}$.insert($\{I^{W}_{j_{\text{min}}} : I^{a}_i\}$)
    \ENDWHILE
\ENDFOR
\STATE $\mathcal{I'}\gets$ ReorderInstrs($\mathbf{Asg}$)
\end{algorithmic}
\end{algorithm}

%% file: sections/op_partition.tex
\section{Operator Partition Pass}
\label{sec:operator_partition}
With the scheduled instructions from the weight gradient computation scheduling pass, we next hide all-to-alls in the forward pass through extensive operator partitioning.
\subsection{Partition Range Selection}
\label{sec:pipeline_range_selection}

Selecting a proper range of non-MoE computation to partition is crucial to maximize overlap and minimize overheads.
Different gating functions affect the type of non-MoE operators we can partition (only ops after the MoE layer, or both before and after the MoE layer).
The number of partitions (how many parts each operator is partitioned into) also affects model performance. 
We introduce a dynamic programming-based algorithm to optimize the aforementioned decisions.

Given an instruction sequence (for forward pass) $\mathcal{I}=[I_1,I_2,\cdots,I_N]$, let $T(n)$ denote the end-to-end execution time (after considering overlapping) of instructions $1$ to $n$ when we have optimally partitioned these instructions.
Then, we have
$$T(n)=\min_{1\leq i \leq n-1}\{T(i) + \min_{1\leq k \leq K}P(i, n, k)\}$$
where $P(i, n, k)$ is the end-to-end execution time of instructions $i$ to $n$ if they are partitioned into $k$ parts and their computation and communication components (all-to-all) arranged to overlap with each other. $K$ is the maximum allowed number of partitions which is a hyper-parameter.
If there is no valid way to partition instructions $i$ to $n$ (e.g., if unsupported gating function is included), we let $P(i,n,k)=\infty$.
The partition axes (dimensions through which the input and output tensors will be split) of each instruction $i,\ldots, n$ is determined by our \textit{partition axis inferencer}.
Then the partitioned instructions are scheduled to overlap each other (form a computation-communication pipeline) by our \textit{pipeline scheduler}, which reports the end-to-end execution time after partition as $P(i,n,k)$.
The optimal end-to-end execution of the entire forward pass of the model is thus $T(N)$.

The dynamic programming algorithm requires $O(N^2 K)$ evaluations of cost $P(i,n,k)$ in total, since there are $N$ $T(n)$s to evaluate and each $T(n)$ requires $K(n-1)$ evaluations of $P(i,n, k)$.
In practice, the number of partitions $k$ is limited by the size of the partitioned dimension (e.g., if we are partitioning along the batch dimension and the batch size is 4, we can have at most 4 partitions). Partition overhead also limits very fine-grained partitioning (in our experiments, we never observed the optimal number of partitions exceeding 4). Therefore, $K$ can be safely set to a relatively small value (e.g., 8).
To further reduce optimization time, we group several consecutive instructions together based on execution time (e.g., total execution time sum up to 2ms) and perform dynamic programming on these groups instead.
Due to partition overhead, an optimal communication-computation pipeline would likely be not very long.
Therefore we can also limit the range of $i$ (i.e., set a maximum length limit on the partition ranges).
Suppose there are $N'$ instruction groups in total and the maximum partition range is $G$ groups. Then the algorithm requires an $O(N'GK)$ number of $P(i,n,k)$ evaluations in total.

\begin{figure}[t]
     \centering
    \begin{subfigure}[]{0.48\columnwidth}
         \centering
         \includegraphics[width=\linewidth]{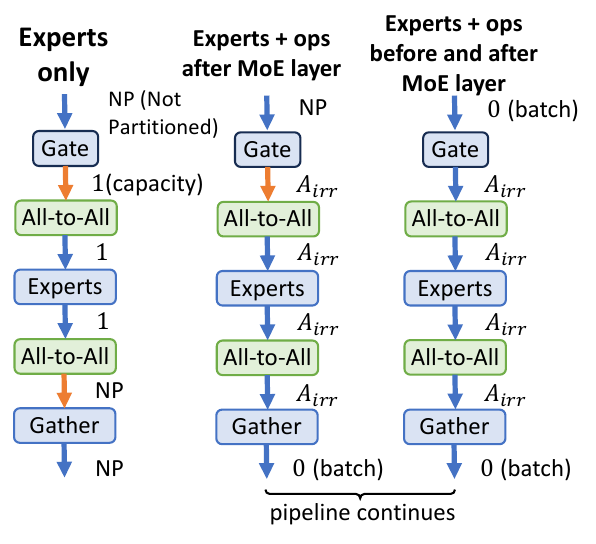}
         \vspace{-5mm}
         \caption{Partition axis of data tensors in different partition types. Orange arrow indicates pipeline begin and end locations, where extra partition/reconstruction instructions are needed.}
         \label{fig:part_ranges_axes}
     \end{subfigure}
     \hfill
     \begin{subfigure}[]{0.48\columnwidth}
         \centering
         \includegraphics[width=\linewidth]{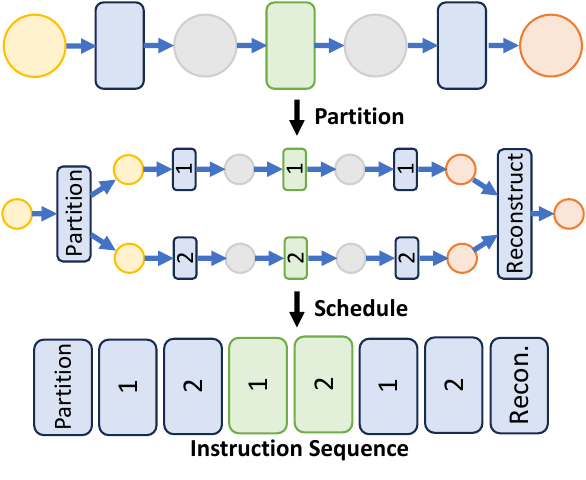}
         \vspace{-5mm}
         \caption{Transforming an instruction sequence to form a pipeline. Yellow (orange) circles denote input (output) tensors. Blue rectangle: computation instruction; Green rectangle: communication.}
         \label{fig:pipeline_gen_steps}
     \end{subfigure}
    \vspace{-3mm}
    \caption{Operator partitioning. }
    \label{fig:pipeline_generation}
\end{figure}

\subsection{Partition Axis Inference}
\label{sec:partition_axis_inference}

To identify the partition axis of each instruction's input and output,
we formulate 
a constraint satisfaction problem.
For the $n$th instruction in the input sequence, let $a^n_{x_i}$ represent the partition axis of its $i$th input, and $a^n_{y_i}$ for its $i$th output.
For each different operator ($f$ in the instructions), we define function $F^f_{\mathcal{Z}}: \mathbf{a_x} \times \mathbf{a_y} \mapsto \mathcal{Z}$ which takes the input and output axes of an instruction $I$ as input and returns a constraint $\mathcal{Z^I}$ (a boolean expression).
Such a constraint specifies how the input and output axes of the instruction should relate for a valid partition (i.e., the original output can be reconstructed from the partitioned ones).
Take matrix multiplication $Y=X\cdot W$ as an example: we can split $X$ along the row ($1$st) dimension and not change $W$, resulting in $Y$ partitioned in the row axis ($\bigl[\begin{smallmatrix} X_1 \\ X_2 \end{smallmatrix}\bigr] W = \bigl[\begin{smallmatrix} X_1 W \\ X_2 W \end{smallmatrix}\bigr]$);
or we can keep $X$ and split $W$ along the column ($2$nd) axis, partitioning $Y$ in the column axis ($X [W_1, W_2] = [XW_1, XW_2]$).
To capture the above possible partition axes combinations, we have the following constraint:
$$(a_{x_1} = 0 \land a_{x_2} = -1 \land a_{y_1} = 0) \lor (a_{x_1} = -1 \land a_{x_2} = 1 \land a_{y_1} = 1)$$
$a_{x_1}$, $a_{x_2}$, $a_{y_1}$ are partition axes for $X$ (the 1st input), $W$ (the 2nd input) and $Y$ (the 1st output) respectively (dimension index starting from 0; -1 means not partitioned).
We also introduce a special partition axis $A_{irr}$ for each MoE-related operator, to represent the irregular partition of all-to-all and experts in our extended computation-communication pipeline (Fig.~\ref{fig:proposed_partition}). The constraints for all-to-alls and experts are written to accept partition at capacity axis if the 
partition range $(i,n)$ only covers the all-to-all and experts, and $A_{irr}$, otherwise.
Correspondingly, we write $F_{\mathcal{Z}}$ of the MoE gather operator to only allow its input to be partitioned at $A_{irr}$ but not the capacity axis, and $F_{\mathcal{Z}}$ of the gating function (if it can be partitioned) to allow batch-partitioned inputs and generate $A_{irr}$ partitioned outputs (Fig.~\ref{fig:part_ranges_axes}).

If the constraints of all instructions are satisfied, every original tensor can be reconstructed from the partitioned ones, asserting correctness of the partition.
We also require that the partition axes of the same tensor cannot be changed, since switching the partition axes requires data from other partitions thus interrupting the computation-communication pipeline.
Putting the above together, we have the following constraint satisfaction problem:

\vspace{-6mm}
\begin{align*}
\textrm{find} \quad & \mathbf{a} \\
\textrm{s.t.} \quad & {F^{f^i}_\mathcal{Z}(\mathbf{a^i_x}, \mathbf{a^i_y}) = 1}, \quad\forall\ i\in [1,N] \\
  & a^i_{y_j} = a^{k}_{x_l}, \ \quad\quad\quad\forall\ (i, j, k, l) \in \mathcal{D}
\end{align*}
where $\mathcal{D}$ describes tensor dependency between operators: indices $(i,j,k,l)\in \mathcal{D}$ if the $j$th output of instruction $i$ is fed to the $l$th input of instruction $k$.

Solving the problem (e.g., using an off-the-shelf solver like OR-Tools~\cite{ortools}) gives the partition axes of all input/output tensors of the partitioned operators. The corresponding partitioned instructions are generated and sent to the pipeline scheduler (Fig.~\ref{fig:pipeline_gen_steps}).

\subsection{Pipeline Scheduling}
\label{sec:pipeline_schedule}

\begin{figure}[t]
    \centering
    \includegraphics[width=0.7\columnwidth]{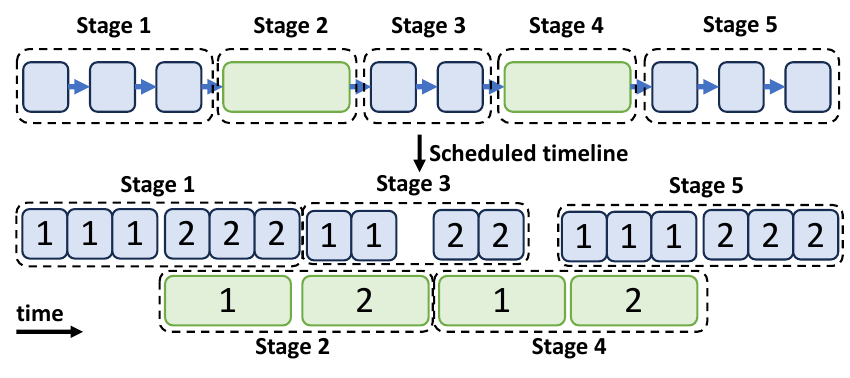}
    \vspace{-5mm}
    \caption{Pipeline schedule by stages. Blue rectangle: computation instructions; Green: communication. Numbers indicate partition index (i.e., the $n$th partition). 
    }
    \label{fig:pipeline_schedule}
    \vspace{-14pt}
\end{figure}

To organize the partitioned instructions into a computation-communication pipeline, the instructions in each partition are divided into stages. Each stage contains all computation or communication that can be consecutively executed.
Instructions in each stage of a partition are always scheduled together.
Within each stage, instructions from the different partition are ordered by partition index (e.g., the first partition always gets scheduled first, and then the second and so on). 
The resulting schedule is demonstrated in Fig.~\ref{fig:pipeline_schedule}.

To obtain the end-to-end (pipelined) execution time of partitioned operators $P(i,n,k)$, we simulate the execution timeline by calculating the start and end time of each instruction relative to pipeline start. Specifically, each instruction's start time is the maximum over (i) the end time of all instructions that it depends on and (ii) the end time of the previous computation/communication instruction (of the same type) in the scheduled order. $P(i,n,k)$ is thus the end time of the last instruction, which is reported back to guide the dynamic programming procedure.

%% file: sections/implementation.tex
\section{Implementation}
\label{sec:implementation}

\SystemName{} is generally applicable to any deep learning compiler for training.
We adopt \CompilerNameWithArticle{}~\cite{yu2023raf}, an open-source compiler extended from Apache TVM~\cite{TVM}, as our underlying compiler, which provides a comprehensive compilation of DL models.
We implement \SystemName{} with 13K LoC in C++.
Communication primitives such as all-to-all are implemented based on NCCL~\cite{nccl}.
\SystemName{} also implements partition constraints ($F_{\mathcal{Z}}$) for all computation operators in common Transformer-based models.
The MoE dispatching ops are implemented based on Tutel's~\cite{hwang2023tutel} kernel. 

Since \SystemName{} is fully implemented in two optimization passes as IR transformations, users only need to enable them in \CompilerName{}'s optimization pass manager, without any modification to the existing code-base. The three hyper-parameters for speeding up the optimization process (i.e., $\rho$, the maximum number of partitions; $\gamma$, the group size; $\iota$, maximum partition range in dynamic programming) can be set through environment variables.

\begin{figure}[t]
     \centering
     \includegraphics[width=\linewidth]{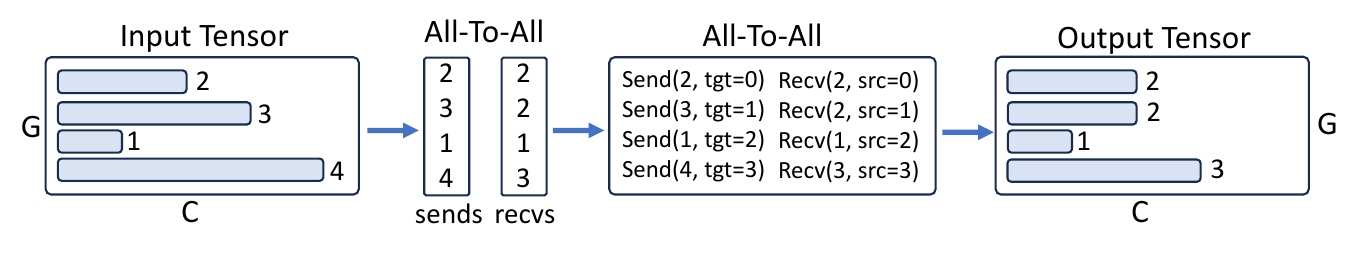}
     \vspace{-8mm}
    \caption{Implementation of irregular all-to-all. G: number of GPUs participating in the all-to-all, $G = E / E_l$ ($E_l$: the number of experts per GPU). On each device, an input and output buffer of fixed shape (G $\times$ C) is allocated. Number in the Input/Output Tensors indicate the actual size of the data to be sent/received on the GPU. The first All-to-All communicates the data sizes to be exchanged; the second All-to-All communicates the actual data. Send/Recv(\textbf{x}, tgt/src=\textbf{y}) indicates an NCCL send/recv primitive that sends/receives a data chunk of size \textbf{x} to/from \textbf{y}.}
    \label{fig:irregular_implementations}
\end{figure}

\noindent\textbf{Irregular all-to-all} (all-to-allv in MPI~\cite{mpi40} terminology) sends different amounts of data to different target devices.
In MoE layers, the amount of data to send to each device depends on the gating function and is only known at runtime (Fig.~\ref{fig:proposed_partition}).
To implement such dynamic communication scheme in a static-shaped system like \SystemName{}, we allocate the input and output tensors based on the maximum amount of data to be sent (i.e., capacity of each expert).
As shown in Fig.~\ref{fig:irregular_implementations}, at runtime, the input buffer is only partially filled based on the result of the gating function.
A first all-to-all is performed to exchange the amount of data to be sent and received across devices, followed by a second all-to-all only sending and receiving the required amount of data.
The all-to-alls are implemented via a grouped NCCL communication consisting of NCCLSends and NCCLRecvs.

%% file: sections/evaluation.tex
\section{Evaluation}

\paragraph{Experiment Setup} We evaluate \SystemName{} on an Amazon EC2 \texttt{p4de.24xlarge} cluster and a \texttt{p3dn.24xlarge} cluster, each with 8 nodes.
Each \texttt{p4de} node has 8 NVIDIA A100 80GB GPUs and 4x100 Gbps NICs.
Each \texttt{p3dn} node has 8 NVIDIA V100 GPUs and one 100 Gbps NIC.
We refer to the cluster of \texttt{p4de.24xlarge} and \texttt{p3dn.24xlarge} nodes as \texttt{A100} and \texttt{V100} respectively, for the rest of the paper.
All nodes run in the same docker environment where we used Ubuntu 20.06 with CUDA 11.3 and NCCL 2.12.12 with PXN enabled.

\paragraph{Benchmark Models and Datasets}
\label{para:benmark_details}
We conduct our evaluations on MoE versions of the GPT-2~\cite{Radford2019GPT2} model (from Huggingface \texttt{transformers}~\cite{wolf-etal-2020-transformers} version 4.18.0). 
The base models are enhanced by replacing every other Transformer block's feed-forward layer with an MoE layer.
Two variants of the model are used: the smaller model (\texttt{GPT2-S-MoE}) has 12 layers with hidden dimension size 768; the larger one (\texttt{GPT2-L-MoE}) has 24 layers with hidden size 1024.
In all experiments, we scale the number of experts along with the number of GPUs: each GPU always hosts two experts.
The SGD optimizer (with momentum) is used for training the model.

For all experiments, we use the WikiText~\cite{merity2016wikitext} dataset as model inputs.
We fix the input sequence length to 512 and use the largest batch size that can fit into the GPU memory for each model: on \texttt{A100}, we use batch size 24 per GPU for \texttt{GPT2-S-MoE} and 48 for \texttt{GPT2-L-MoE}.
On \texttt{V100}, we use batch size 16 for \texttt{GPT2-S-MoE} and 8 for \texttt{GPT2-L-MoE}.

\begin{figure*}[ht]
    \centering
     \begin{subfigure}[b]{0.24 \linewidth}
     \centering
     \includegraphics[width=\linewidth]{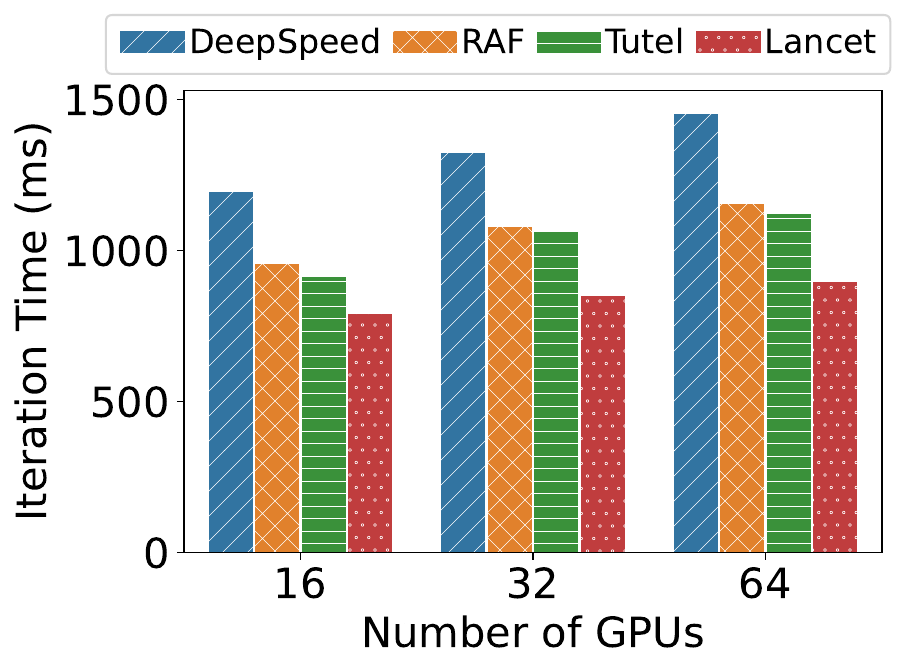}
     \vspace{-5mm}
     \caption{GPT2-S-MoE, \texttt{V100}.}
     \label{fig:throughput_gpt2_s_p3dn_switch}
    \end{subfigure}
     \hfill
     \begin{subfigure}[b]{0.24 \linewidth}
         \centering
         \includegraphics[width=\linewidth]{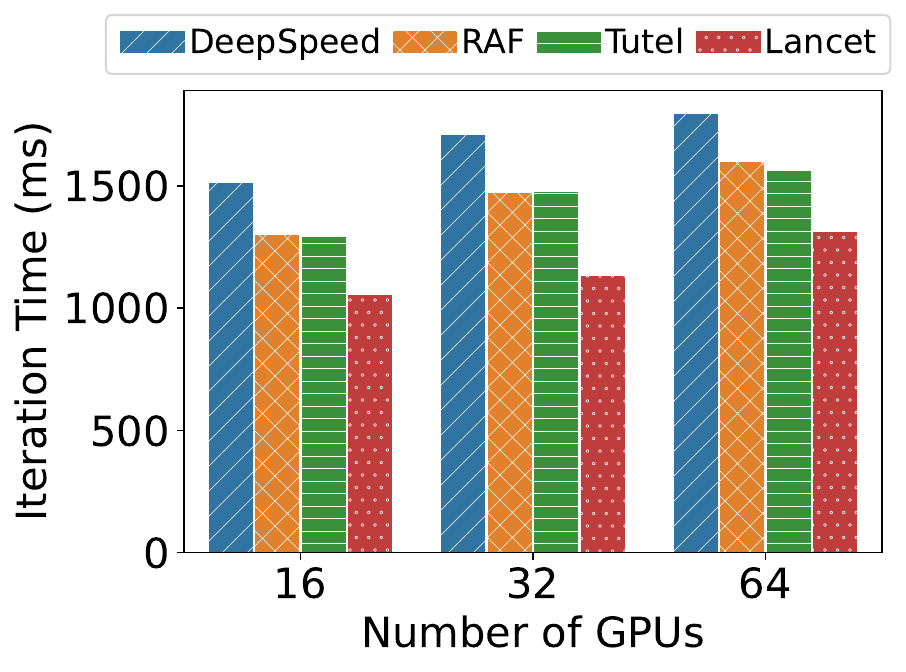}
         \vspace{-5mm}
         \caption{GPT2-L-MoE, \texttt{V100}.}
         \label{fig:throughput_gpt2_l_p3dn_switch}
     \end{subfigure}
     \hfill
     \begin{subfigure}[b]{0.24 \linewidth}
         \centering
         \includegraphics[width=\linewidth]{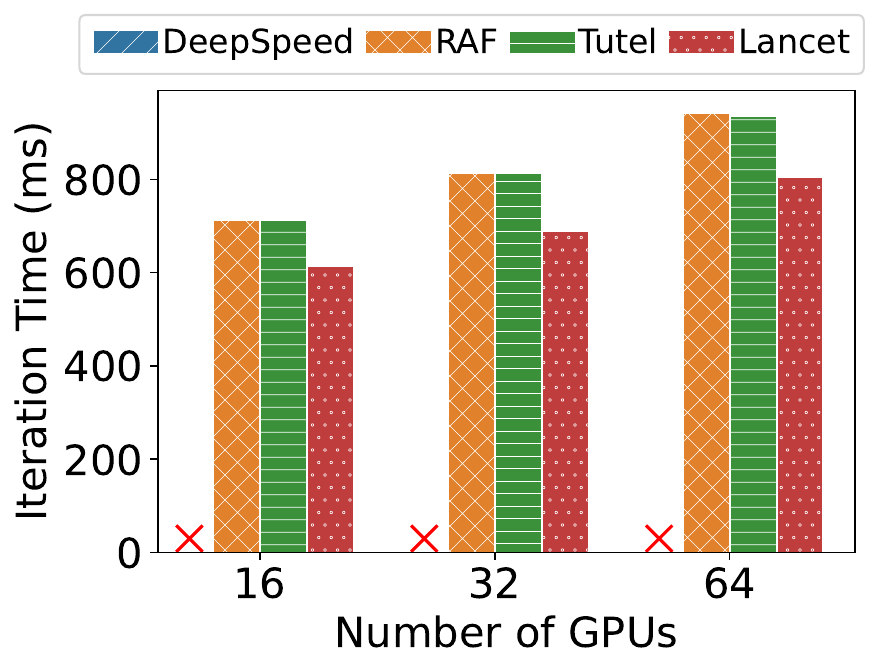}
         \vspace{-5mm}
         \caption{GPT2-S-MoE, \texttt{A100}.}
         \label{fig:throughput_gpt2_s_p4de_switch}
     \end{subfigure}
      \hfill
     \begin{subfigure}[b]{0.24 \linewidth}
         \centering
         \includegraphics[width=\linewidth]{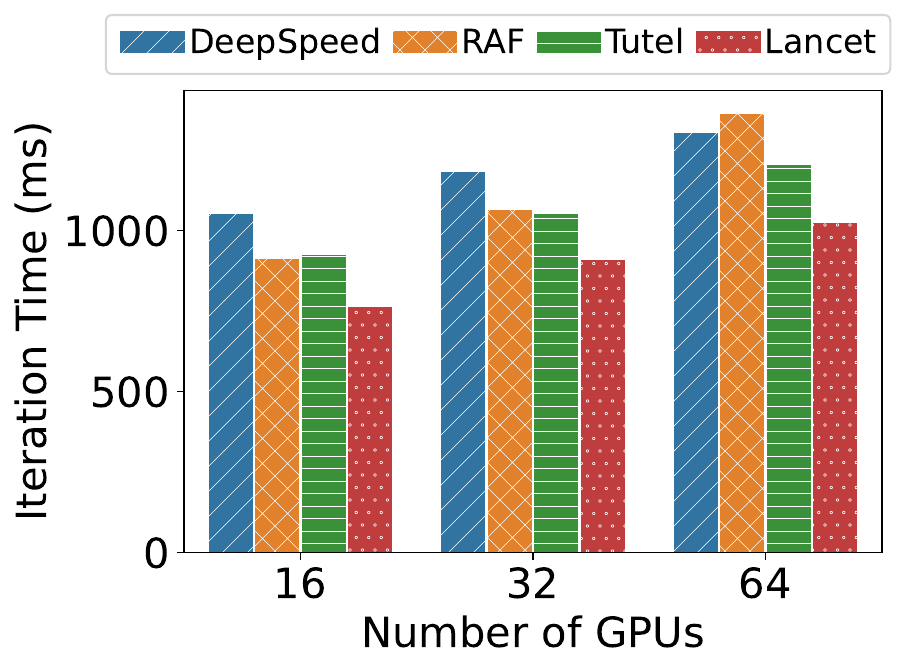}
         \vspace{-5mm}
         \caption{GPT2-L-MoE, \texttt{A100}.}
         \label{fig:throughput_gpt2_l_p4de_switch}
     \end{subfigure}
    \vspace{-3mm}
    \caption{Training iteration time when using Switch gate. Red cross indicates out-of-memory.}
    \label{fig:throughput_scaling_switch}
    \vspace{-3mm}
\end{figure*}

\begin{figure*}[ht]
    \centering
     \begin{subfigure}[b]{0.24 \linewidth}
     \centering
     \includegraphics[width=\linewidth]{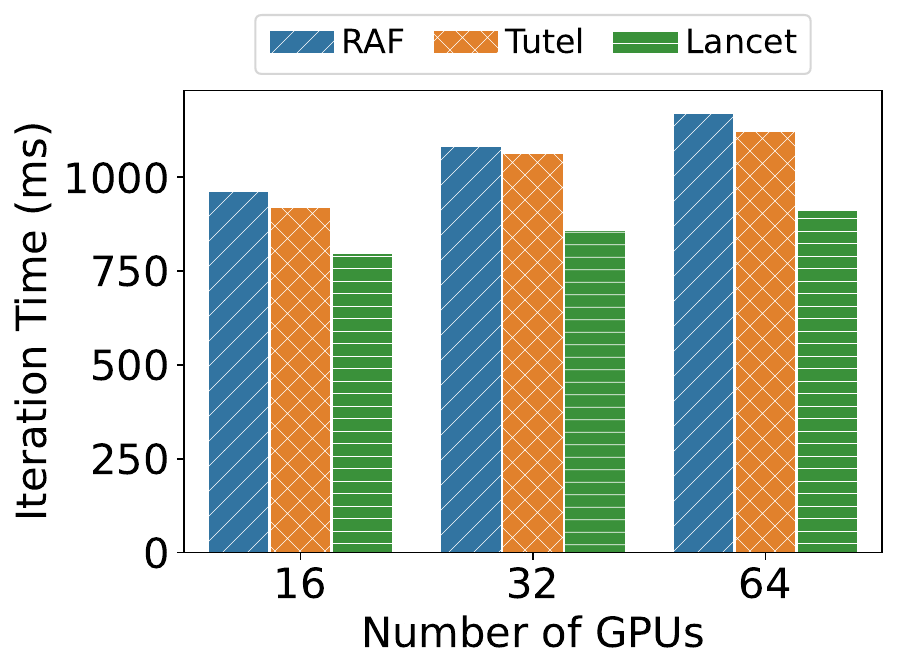}
     \vspace{-5mm}
     \caption{GPT2-S-MoE, \texttt{V100}.}
     \label{fig:throughput_gpt2_s_p3dn_bpr}
    \end{subfigure}
     \hfill
     \begin{subfigure}[b]{0.24 \linewidth}
         \centering
         \includegraphics[width=\linewidth]{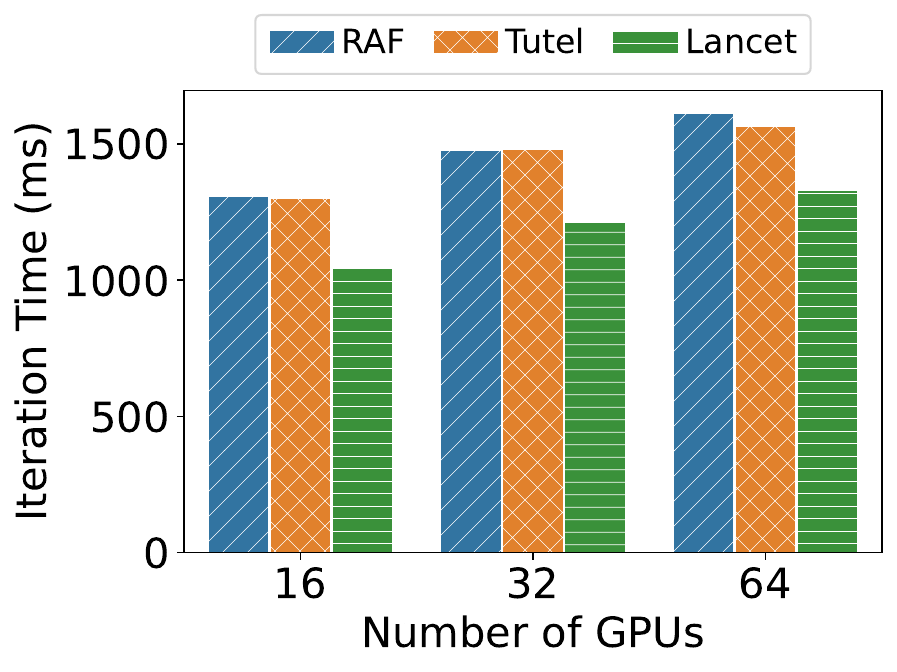}
         \vspace{-5mm}
         \caption{GPT2-L-MoE, \texttt{V100}.}
         \label{fig:throughput_gpt2_l_p3dn_bpr}
     \end{subfigure}
     \hfill
     \begin{subfigure}[b]{0.24 \linewidth}
         \centering
         \includegraphics[width=\linewidth]{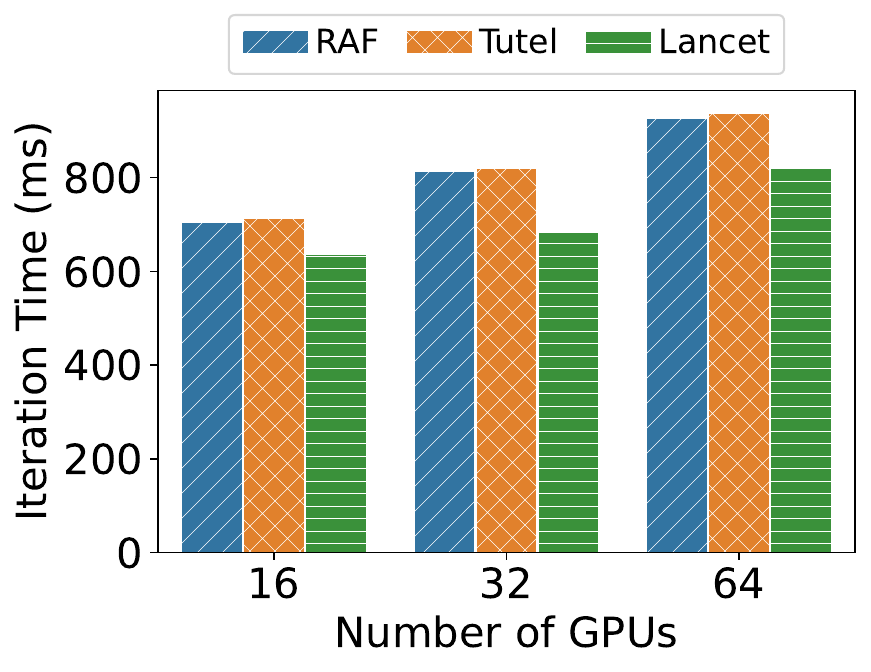}
         \vspace{-5mm}
         \caption{GPT2-S-MoE, \texttt{A100}.}
         \label{fig:throughput_gpt2_s_p4de_bpr}
     \end{subfigure}
      \hfill
     \begin{subfigure}[b]{0.24 \linewidth}
         \centering
         \includegraphics[width=\linewidth]{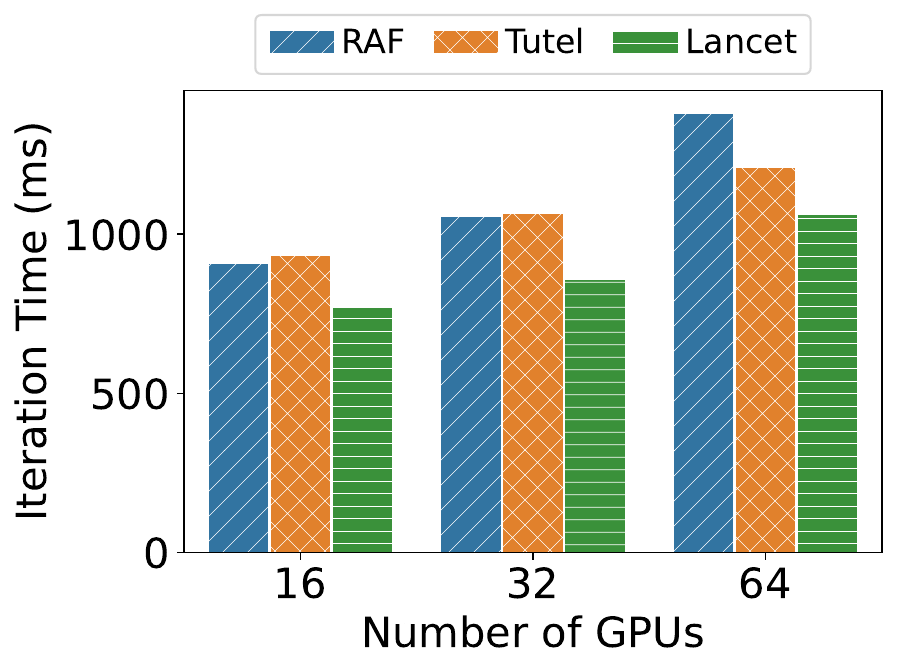}
         \vspace{-5mm}
         \caption{GPT2-L-MoE, \texttt{A100}.}
         \label{fig:throughput_gpt2_l_p4de_bpr}
     \end{subfigure}
    \vspace{-3mm}
    \caption{Training iteration time when using Batch Prioritized gate.}
    \label{fig:throughput_scaling_bpr}
    \vspace{-3mm}
\end{figure*}

\paragraph{Baselines}
We compare \SystemName{}'s training performance with DeepSpeed (version 0.5.8, without Tutel's kernels)~\cite{rasley2020deepspeed} and Tutel (version 0.3)~\cite{hwang2023tutel}.
Tutel implements overlapping between all-to-all and expert computation.
For each experiment with Tutel, we search through the overlapping degree (the number of partitions) of 1, 2, 4 and 8 and report the best result.
Tutel and DeepSpeed are both built on PyTorch~\cite{adam2019pytorch},
whose performance on computation ops may be different from RAF~\cite{yu2023raf}. 
Therefore, we also include results of RAF without \SystemName{}'s modifications for comparison.

\paragraph{Hyper Parameters}
We set the maximum number of partitions $\rho$ to 8, except when excessive partitions cause out-of-memory (OOM) errors.
In that case, we reduce it to 4 (and 2 if still OOMs).
We set the group size $\gamma$ according to the model execution time so that there are 5 groups between each MoE layer.
The maximum partition range $\iota$ is set to be the execution time between two MoE layers, so one pipeline will be formed per MoE layer.

\subsection{Throughput}
We compare \SystemName{}'s training throughput against baselines using different numbers of GPUs.
We do weak scaling, i.e., keep the local batch size fixed at each GPU while the effective total batch size of the model scales linearly.
Since gating method constraints the available pipeline range, we run the experiments with two different gating methods: Switch~\cite{fedus2022switch} gate which allows overlapping with computation both before and after the MoE layer (Fig.~\ref{fig:part_all}) and Batch Prioritized~\cite{riquelme2021bpr} gate which only allows overlapping with computation after the MoE layer (Fig.~\ref{fig:part_after_gate}).

Fig.~\ref{fig:throughput_scaling_switch} shows that \SystemName{} achieves up to 1.21x (1.17x on average) speed up compared to the baselines on the \texttt{A100} cluster, and up to 1.3x (1.22x on average) on \texttt{V100} cluster when using Switch gate.
We find DeepSpeed exhibits slightly higher memory requirements than other frameworks, leading to OOM on \texttt{A100} when running the GPT2-S-MoE model (OOM does not happen on \texttt{V100} since a smaller batch size is used, i.e., 24 v.s. 16).
When using Batch Prioritized gate (Fig.~\ref{fig:throughput_scaling_bpr}), we observed up to 1.24x (1.17x on average) speed up on the \texttt{A100} cluster, and up to 1.24x (1.21x on average) on \texttt{V100} cluster.
Despite more constraint pipeline range, the achieved speed up for Batch Prioritized gate is overall similar to that of the Switch gate.
This is because despite only pipelining with computation after the MoE layer, significant amount of overlapping can still happen.
Our $dW$ scheduling is also unaffected by the gating methods.
The maximum achieved speed up on \texttt{V100} is lower when using Batch Prioritized gate though, indicating that partitioning may have a larger impact on \texttt{V100}.

\begin{figure}[t]
    \centering
     \begin{subfigure}[b]{\columnwidth}
     \centering
     \includegraphics[width=\linewidth]{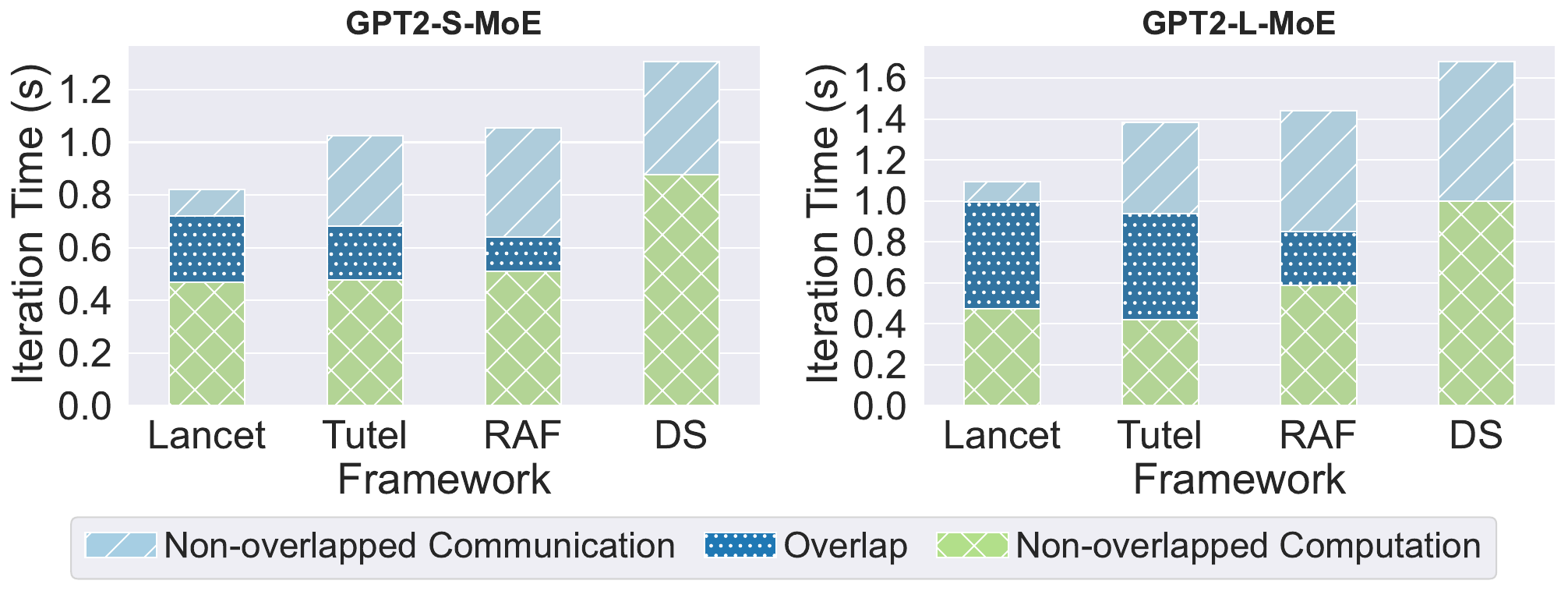}
     \vspace{-4mm}
     \caption{On 4 \texttt{V100} nodes}
     \label{fig:time_decompose_p3dn}
    \end{subfigure}
     \hfill
     \begin{subfigure}[b]{\columnwidth}
         \centering
         \includegraphics[width=\linewidth]{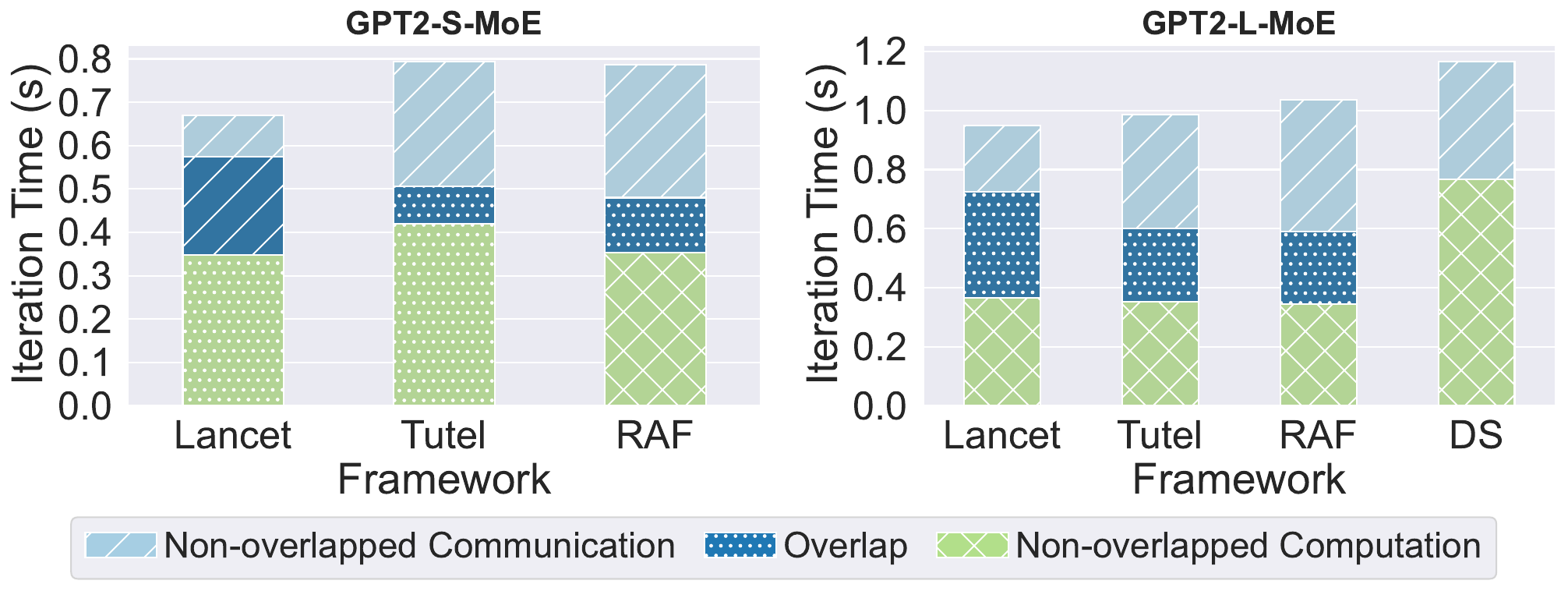}
         \vspace{-4mm}
         \caption{On 4 \texttt{A100} nodes.}
         \label{fig:time_decompose_p4de}
     \end{subfigure}
    \vspace{-8mm}
    \caption{Iteration time decomposition. DS: DeepSpeed.
    }
    \label{fig:time_decomposition}
\end{figure}

As shown in Fig.~\ref{fig:time_decomposition}, \SystemName{} achieves a higher level of computation-communication overlapping than baselines, reducing non-overlapped communication time by up to 69\% (\texttt{A100}) and 83\% (\texttt{V100}) compared to RAF, 66\% (\texttt{A100}) and 77\% (\texttt{V100}) compared to Tutel.
The trade-off of applying partition-pipeline is also clearly shown in Fig.~\ref{fig:time_decomposition}.
While \SystemName{}'s optimizations decrease the end-to-end execution time, the total execution time of computation (Non-overlapped Computation + Overlapped) ops can be higher than that of RAF, due to partition overheads. Since \SystemName{} implements irregular all-to-alls and do not transmit any padding tokens between experts, the overall communication time (Non-overlapped Communication + Overlapped) can be lower than baselines.

\subsection{Accuracy of cost model}
\label{sec:accuracy}

\begin{figure}[t]
    \vspace{-3mm}
     \centering
     \includegraphics[width=0.8\linewidth]{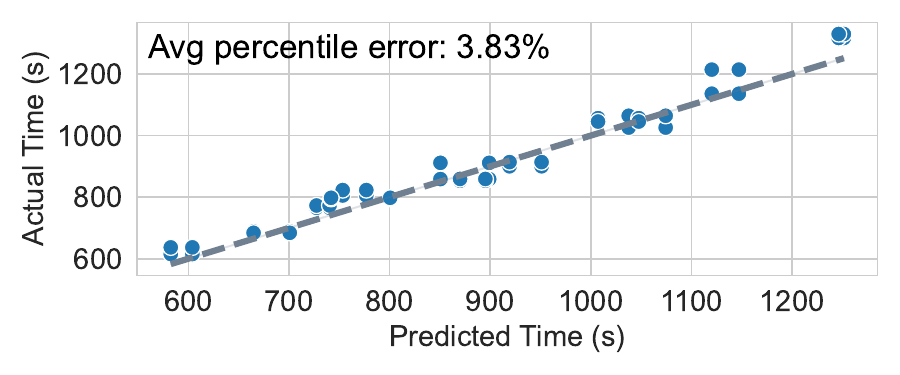}
     \vspace{-5mm}
     \caption{Prediction accuracy of \SystemName{}'s cost model. Data aggregated from all models bench-marked on all clusters during our experiments.
     }
     \label{fig:preduction_accuracy}
     \vspace{-5mm}
\end{figure}

Fig.~\ref{fig:preduction_accuracy} shows the accuracy of \SystemName{}'s cost model, used to predict the iteration time after applying each optimization.
The prediction error is very small (3.83\%).
Such an accurate cost model provides useful information to guide our weight gradient computation scheduling and DP-based operator partitioning algorithms.

\subsection{Optimization Time}
\begin{figure}[t]
    \centering
     \begin{subfigure}[b]{0.49\columnwidth}
     \centering
     \includegraphics[width=\linewidth]{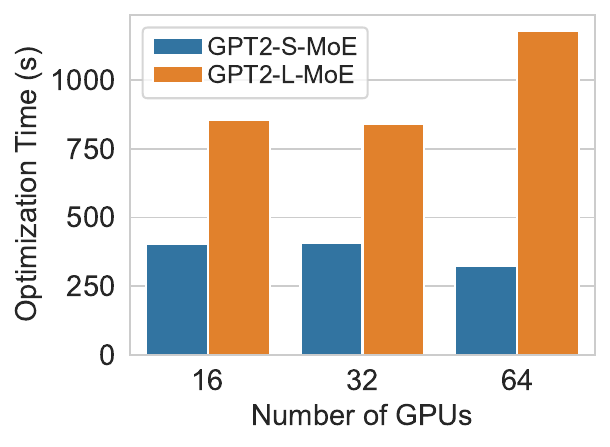}
     \vspace{-7mm}
     \caption{\texttt{V100} cluster}
     \label{fig:opt_time_v100}
    \end{subfigure}
    \hfill
     \begin{subfigure}[b]{0.49\columnwidth}
     \centering
     \includegraphics[width=\linewidth]{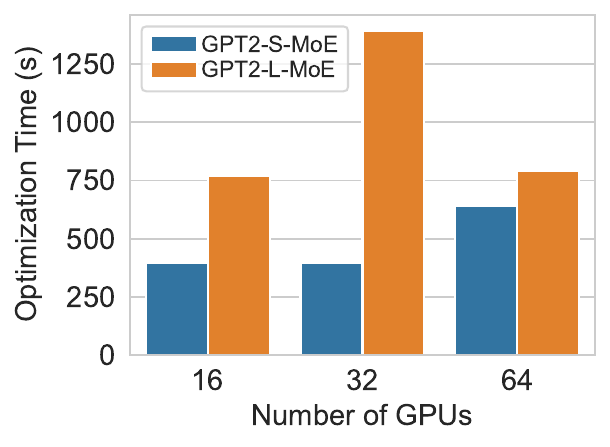}
     \vspace{-7mm}
     \caption{\texttt{A100} cluster}
     \label{fig:opt_time_a100}
    \end{subfigure}
    \vspace{-3mm}
    \caption{\SystemName{}'s optimization time when using Switch gate.
    }
    \label{fig:optimization_time}
    \vspace{-5mm}
\end{figure}

Fig.~\ref{fig:optimization_time} shows the time taken to optimize the models in our experiments. Optimization time is dominated by the operator partition pass (Sec.~\ref{sec:operator_partition}) since weight gradient computation schedule (Sec.~\ref{sec:dw_schedule_pass}) uses a fast greedy algorithm.
Since every device shares the same computation graph, the optimization time is less affected by the number of GPUs used and more by the number of layers in the model.
The optimization time of most models bench-marked is below 20 minutes.
Our optimization also only requires one GPU to run (for bench-marking execution time of partitioned computation ops).

\subsection{Ablation Study}
\begin{figure}[t]
    \centering
     \begin{subfigure}[b]{\linewidth}
     \centering
     \includegraphics[width=0.9\linewidth]{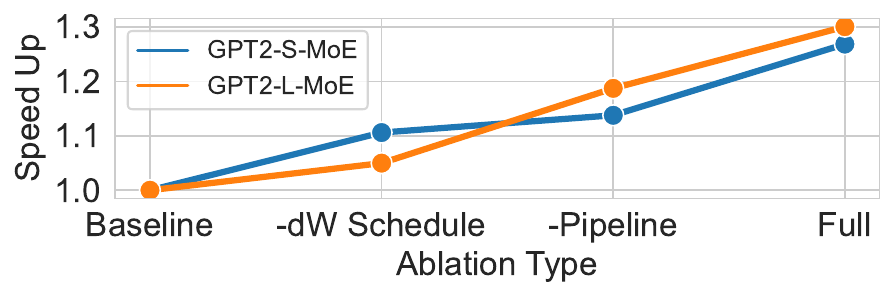}
     \vspace{-3mm}
     \caption{on 4 \texttt{V100} nodes}
     \label{fig:ablation_p4d}
    \end{subfigure}
    \hfill
    \begin{subfigure}[b]{\linewidth}
     \centering
     \includegraphics[width=0.9\linewidth]{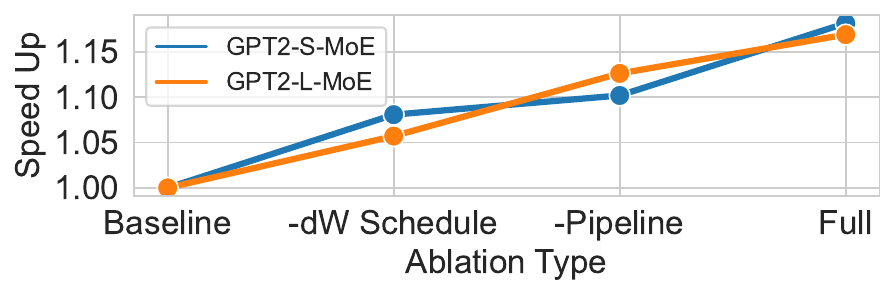}
     \vspace{-3mm}
     \caption{on 4 \texttt{A100} nodes}
     \label{fig:ablation_p3dn}
    \end{subfigure}
    \vspace{-7mm}
    \caption{Ablation study on 4 \texttt{A100} and \texttt{V100} nodes. $dW$: weight gradient computation.}
    \label{fig:ablation_study}
\end{figure}

To show the effects of weight gradient computation scheduling and pipelining separately, we conduct an ablation study on 4 \texttt{A100} and \texttt{V100} nodes.
In Fig.~\ref{fig:ablation_study}, the relative speed-up is computed by dividing the training throughput under each scheme by that of RAF without any \SystemName{} optimizations.
For both models, applying only scheduling or only pipelining yields a lower speedup compared to using them together.
On both clusters, \texttt{GPT2-L-MoE} is affected more by disabling weight gradient computation scheduling, while the two optimizations have more similar performance gain on \texttt{GPT2-S-MoE}.
This is because \texttt{GPT2-L-MoE} has more parameters and layers while using a smaller batch size, thus having higher partition overheads, rendering weight gradient computation scheduling more effective compared to operator partition.

\section{Discussion and Related Works}

\paragraph{Compatibility with other large-scale training techniques}
While \SystemName{} is evaluated with data and expert parallelism, the techniques are in principle compatible with most other commonly used training optimizations. Weight gradient scheduling only utilizes operator dependency during backward propagation, thus unaffected by most distributed training sharding techniques. Some techniques introduce extra communication which may interfere with partition-based all-to-all overlapping. FSDP/ZeRO3~\cite{rajbhandari2020zero} inserts additional all-gather communication in the forward passes, which may require additional scheduling to avoid interference with overlapped all-to-all. Tensor parallelism~\cite{shoeybi2019megatron} requires all-reduce communication after self-attention; Ring-attention (sequence parallelism) ~\cite{liu2023ring} communicates the key-value blocks during the attention process. If different devices or communication channels are used for expert and tensor/sequence parallelism (e.g., inter-node vs. intra-node), the overlapped all-to-all communication can be arranged to execute concurrently with tensor/sequence parallelism traffic. Investigating the efficient orchestration and overlapping of communication arising from various sharding techniques, particularly the intricate patterns generated by automatic sharding~\cite{zheng2022alpa}, remains future work.

\paragraph{Optimizing irregular communication and expert computation}
\SystemName{}'s partition produces irregular-shaped all-to-alls and expert computation. While we use a simple NCCL based implementation (Fig.~\ref{fig:irregular_implementations}), better communication implementations targeting such dynamic workload may further improve the performance. Similarly, the shape irregularity in expert computation may cause extra computation due to padding. Block-sparse expert kernels (e.g., MegaBlocks~\cite{megablocks}) can be further applied to accelerate the computation. 

\paragraph{MoE architectures that facilitate overlapping}
PR-MoE~\cite{rajbhandari2022deepspeed-moe} and DeepSeek-MoE~\cite{dai2024deepseekmoe} use a shared expert which all tokens are routed to. The all-to-all communication (for non-shared experts) can also be overlapped with the computation of such shared expert. \SystemName{}'s approach can be applied to a wider-range of MoE models that use traditional architectures, e.g., ~\cite{jiang2024mixtral}.

\paragraph{Other MoE training optimization techniques}
Tutel~\cite{hwang2023tutel} and FasterMoE~\cite{he2022fastermoe} are two popular frameworks optimizing for MoE models.
Both frameworks support overlapping all-to-all and expert computation.
Tutel~\cite{hwang2023tutel} also implements fast dispatching kernels, better all-to-all algorithm, and adaptive parallelism switching for dynamic workloads. FasterMoE~\cite{he2022fastermoe} proposes techniques to handle imbalanced expert selection and to select experts based on network topology.
These optimizations are orthogonal to ours and can potentially be used in conjunction.
\cite{zhang2022accelerating} proposes to run two copies of the model on the same device, overlapping computation and communication between different model replicas. However, splitting the input among the two model replicas may result in mathematical in-equivalence (e.g., due to extra token dropping).
\cite{li2023lina} optimizes MoE training by prioritizing all-to-all traffic over all-reduce traffic, avoiding bandwidth contention and improving all-to-all latency. This method can also be used in conjunction with \SystemName{}.

\section{Conclusion}
This paper presents \SystemName{}, a system to automatically optimize MoE model training.
We extend the optimization space of current methods and seek whole-training-graph-level opportunities to overlap all-to-all communication. 
In the forward pass, we overlap all-to-all with both expert and non-MoE computation through proper partitioning and pipelining.
The optimal partition range is determined by a dynamic programming algorithm.
In the backward pass, we schedule weight gradient computation to overlap all-to-all using an best-fit greedy algorithm.
Experimental evaluation shows that \SystemName{} reduces non-overlapped communication time by up to 77\%, and achieves up to 1.3x end-to-end speed up compared to state-of-the-art solutions.

\section{Acknowledgements}
We would like to thank the anonymous reviewers for their valuable feedback.
This work was supported by an Amazon Research Award (ARA) on AWS AI and grants from Hong Kong RGC under the contracts HKU 17208920, 17204423 and C7004-22G (CRF).